\documentclass[aps,pre,groupedaddress,twocolumn]{revtex4-2}

\usepackage{amsmath}    
\usepackage{graphicx}   
\usepackage{verbatim}   
\usepackage{color}      
\usepackage{subfigure}  
\usepackage{hyperref}   
\newcommand{\bq}{\begin{equation}}
\newcommand{\eq}{\end{equation}}
\newcommand{\bqn}{\begin{eqnarray}}
\newcommand{\eqn}{\end{eqnarray}}
\newcommand{\nb}{\nonumber}
\newcommand{\lb}{\label}
\newcommand{\q}{\vec{q}}

\newcommand{\rf}{r_{\text{eff}}}
\newcommand{\Ni}{{\cal{N}}^{-1}}
\newcommand{\D}{{\cal{D}}}

\begin{document}


\title{A Data-Driven Statistical Description for the Hydrodynamics of Active Matter}


\author{Ahmad Borzou$^{1}$}

\author{Alison E. Patteson$^1$}
\author{J. M. Schwarz$^{1,2}$}
\affiliation{$^1$Physics Department and BioInspired Institute,Syracuse
  University, Syracuse, NY 13244 $^2$Indian Creek Farm}


\date{\today}

\begin{abstract}
Modeling living systems at the
collective scale can be very challenging because the 
individual constituents can themselves be complex and the respective
interactions between the constituents are not fully
understood. With the advent of high throughput experiments and in the
age of big data, data-driven methods are on the rise to overcome these
challenges. Although
machine-learning approaches can help quantify correlations between the
various players, they do
not directly shed light on the underlying physical principles of
such systems. To directly uncover the underlying physical
principles, we present a data-driven method for obtaining the phase-space
density such that the solution to the stochastic dynamic equation for active
matter readily emerges, from which time and space dependence of physical order parameters can be readily extracted. 
If the system is near a steady state,  we illuminate how to construct a field theory to subsequently make
physical predictions about the system. 
The method is first developed analytically and subsequently calibrated using simulated data. The method is
then applied to an experimental system of particles
actively driven by a {\it Serratia marcescens}
bacterial swarm and in the presence of spatially localized UV light. The analysis demonstrates that the particles are in the steady-state
before and sometime after the UV light and obey a
Gaussian field theory with a spatially-varying ``mass'' in those
regimes. This novel, yet simple, finding is surprising given the
complex dynamics of the bacterial swarm. In response
to the UV light, we demonstrate that there is a net flow of the particles away from
the UV light and that the entropy of
the particles increases away from the light. We conclude with
a discussion of additional potential applications of our data-driven method such as when the internal structure of the individual constituents
dynamically changes to result in a modified stochastic dynamic equation governing the system. 
\end{abstract}


\maketitle

\section{Introduction}
Complexities of living matter include its multi-scale nature, its multi-body nature, its
disrespect of time-reversal symmetry, and its ability to replicate and
to learn, or adapt, from which new properties and/or symmetries may
emerge. An example of a new emergent property at the cellular level is
the elongation of cells during development~\cite{Paluch_2009} and at the onset of
some diseases~\cite{Lamouille_2014}. With such changes comes new, emergent
interactions that need to be quantified. The above complexities call for a focus on {\it in vitro}
systems, with only a few types of players, or constituents, if we are to build
quantitative models of living matter with predictive power.   

One of the candidate frameworks to model living matter is known as
active matter~\cite{Ramaswamy_2010,Vicsek_2012,Marchetti_2013,Gompper_2020}.
In active matter, each constituent is internally driven, typically by
consuming energy to generate motion.  With this construction,
non-living material such as colloids, microbots,  and other
self-propelled particles also fall under the purview of active
matter~\cite{Zottl2016,Lozano2016,Vutukuri2020,Giomi_2013,Chvykov_2021}. To
date, two main categories of theoretical approaches have been used to
quantify active matter: analytical analysis of dynamic stochastic equations of an
assumed form (see, for example, equations governing the hydrodynamics of
flocking~\cite{Toner_Tu_PRL_1998,Toner_Tu_PRE_1998}) and more
intricate computational approaches.  While results from the former 
category are robust, the assumptions involved may
limit its applicability. 

The challenges of the analytic approaches have motivated machine-learning approaches at the opposite end of the spectrum of quantitative frameworks~\cite{Cichos_2020}. 
With machine-learning techniques, one searches out correlations by sifting through reams of data.  
While this approach comes with many advantages, there are shortcomings as well.  
One of the difficulties in machine-learning is the freedom in model
selection, which could contribute towards the non-reproducibility issues reported in \cite{baker20161}. 
Application of machine-learning methods in the exact sciences, such as
physics, encounters the additional problem of violating
basic physical principles. In other words, both feature and model selections in machine-learning are not constrained as severely as
principles in physics. Therefore, a researcher may train a
model that eventually violates the second law of thermodynamics, or
the system's energy may not be bound from below such that ghosts with
negative kinetic energy are allowed to propagate.  We should mention
that there has been very recent progress in dealing with this very issue~\cite{PhysRevLett.126.098302}.

Our approach in this manuscript is to take advantage of learning from
data but remaining within the strict framework of the governing laws of physics.  Therefore, instead of working with the conventional machine-learning
models, we adopt the conventional statistical models of many-body systems to
be trained by data, where the phase-space distribution function serves as the prominent
feature of the model.  An advantage of our approach to learning
from data is that the interpretation of both the model and its features
is well understood.  The approach also aims at overcoming the
challenge of dealing with unknown interactions, emergent or otherwise.  The expectation with this hybrid approach is that new and
old physical principles will readily emerge in living systems at the collective scale. 

We start with the time evolution equation for the phase-space density
function as the underlying equation generating the non-equilibrium
field equations for the order parameters, such as the number density
and the polarization vector.  The dynamical equation of the phase-space density is under-determined in its general form but is exact.  
In analytic approaches, one needs to make simplifying assumptions
and approximations to overcome the latter problem to find analytic
solutions to the field equations.  In our approach, we work with the
exact dynamical equation and estimate the phase-space density directly
from data.  Since in an experiment, the phase-space density function
is truly driven by the exact equation, our estimation would be the
solution to the exact field equations with no assumptions required.   
Also, we show that our data-driven approach can be used to build an
effective statistical field theory for those systems that are close to
steady state. 
For this purpose,  we use our data-driven description of the time
evolution of the order parameters to observe their fluctuations over
time and their correlations over space. Since the latter is related to
the functional derivatives of the effective action, we can then solve
a system of equations to reconstruct an analytic effective action.
Finally, we test our method using both simulations and experiments.

The structure of this paper is as follows. 
We construct the theoretical framework in Sec.~\ref{Sec:Theory}.  We
validate our data-driven method in three simulations that are
presented in Sec.~\ref{Sec:Simulations}.  In Sec.~\ref{Sec:Data},  we apply our method to 
an experimental system of spherical particles embedded within a bacterial swarm that respond to localized UV light. 
We draw conclusions in Sec.~\ref{Sec:Conclusion}.

\section{Theoretical setup}
\lb{Sec:Theory}
The breaking of time-reversal
  symmetry in active matter results in a non-equilibrium system making the direct connection
  with equilibrium statistical mechanics approaches untenable,
  particularly in the absence of any steady-state. 
  However, non-equilibrium statistical mechanics approaches exist.  Generally speaking, the stochastic dynamics of any observable $\hat{A}(\Gamma,t)$ in a statistical system can be written as 
 \bqn
\frac{\partial \hat{A}(\Gamma,t)}{\partial t} = {\cal{L}} \hat{A}(\Gamma,t),
 \eqn
where $\Gamma$ is a point in the N-body phase-space,  and ${\cal{L}}$
consists of combinations of derivatives with respect to real-space
positions,  velocities,  angles,  angular velocities,  etc.  of each
of the N particles in the system.  It should be noted that the exact
form of ${\cal{L}}$ depends on the system of interest.  For
self-propelled rods, for example,  its form is given in
Ref.~\cite{Baskaran_2010},  which is found by starting with the Ito calculus \cite{gardiner2004handbook}. 
 
A measurable quantity is the ensemble average of the observable, which reads
\bqn
\lb{Eq:ADynamic}
\langle \hat{A}(t) \rangle &=& 
\int d\Gamma\, \hat{f}_N(\Gamma)\,  \hat{A}(\Gamma,t),\nb\\
&=& \int d\Gamma\, \hat{f}_N(\Gamma,t)\,  \hat{A}(\Gamma),
\eqn
where $\hat{f}_N(\Gamma)$ is the N-body phase-space probability density, and in the second line, the dynamics is equivalently passed to the latter density function.  Taking partial derivatives of Eq.~\eqref{Eq:ADynamic} and integrating by parts, we conclude that 
\bqn
\lb{Eq:TimeEvol_fN}
\frac{\partial \hat{f}_N(\Gamma,t)}{\partial t} + {\cal{L}} \hat{f}_N(\Gamma,t) = 0. 
\eqn

We define the $m$-body phase-space density function as 
\bqn
\lb{Eq:fmDef}
f_m(\Gamma_1, \cdots, \Gamma_m) \equiv \int d\Gamma_1\,d\Gamma_2\, \cdots\, d\Gamma_m  \hat{f}_N(\Gamma,t). 
\eqn
As will be discussed later, most of the order parameters in active matter,  such as the number density in real-space, the polarization vector, and the nematic tensor, are known in terms of $f_1$.  Therefore, we are interested in the time evolution of the latter quantity.  
In the following,  for simplicity,  we drop the subscript and refer to the one-body distribution function as $f\equiv f_1$.

We apply $\frac{\partial}{\partial t}$ to both sides of Eq.~\eqref{Eq:fmDef} for $m=1$ and use Eq.~\eqref{Eq:TimeEvol_fN} to rearrange the terms in the following form
\bqn
\lb{Eq:BoltzmannEq}
\frac{df}{dt} = \frac{\partial f}{\partial t} + \frac{\partial f}{\partial q^{\mu}}\frac{d q^{\mu}}{dt} = C,
\eqn
where points in the one-body phase-space are labeled by $\vec{q}\equiv
(\vec{x},\vec{v},\hat{u},\vec{\omega}, \cdots)$, consisting of positions,  velocities,  orientations,  angular velocities,  and other possible internal structures,  and repeated indices indicate a sum.   We will discuss $C$ in the following subsections.

\subsection{Hydrodynamic field equations}
\lb{Sec:Hydro}
For simplicity,  we assume that $\vec{q}\equiv
(\vec{x},\vec{v},\hat{u},\vec{\omega})$ and neglect the rest of the possible internal structures of the particles.  
To arrive at the field equations for the order parameters in active matter,  we need to compute the moments of Eq.~\eqref{Eq:BoltzmannEq}. 
The continuity equation is the zeroth moment of Eq.~\eqref{Eq:BoltzmannEq} and reads
\bqn
\lb{Eq:Boltzmann0thMom}
\partial_t \rho + \partial_{\mu}\left(\rho \bar{v}_{\mu}\right) = \int dv\, d\hat{u}\, d\omega\, C,
\eqn
where $\partial_{\mu}\equiv \frac{\partial}{\partial x^{\mu}}$,  $dv$,  $d\hat{u}$,  and $d\omega$ refer to the ``volume'' integrals in their corresponding spaces, and the particle density and the average velocity are defined respectively by 
\bqn
\lb{Eq:densityAndVelocity}
&&\rho\left(\vec{x},t\right)\equiv \int dv\, d\hat{u}\,d\omega\,  f,\nb\\
&& \bar{v}_{\mu}\left(\vec{x},t\right) \equiv \frac{1}{\rho}\int dv\, d\hat{u}\,d\omega\,  f v_{\mu}.
\eqn
Also, we have set the surface term from $\frac{\partial}{\partial \hat{u}_{\mu}}$ equal to zero. 

The time evolution of the average velocity can be found through the first velocity moment of Eq.~\eqref{Eq:BoltzmannEq}
\bqn
\lb{Eq:Boltzmann1stMom}
\partial_t\left(\rho \bar{v}_{\mu}\right)+ \partial_{\nu}P_{\mu\nu} = \int  dv\, d\hat{u}\,d\omega\,  v_{\mu}C,
\eqn
where we have set $\frac{dv^{\mu}}{dt}=\frac{d\omega^{\mu}}{dt}=0$.  Also,  we assumed that the external forces applied to the particles are zero.
Moreover,  the stress tensor is defined by
\bqn
\lb{Eq:StressTensor}
P_{\mu\nu} \equiv \int  dv\, d\hat{u}\,d\omega\,  f v_{\mu} v_{\nu}.
\eqn
It should be noted that the definition of the stress tensor might be different in other sources where some of the terms
from the right-hand side of Eq.~\ref{Eq:Boltzmann1stMom} are absorbed into the definition.  Nevertheless, we prefer our definition of the stress tensor because (i) it is Eq.~\eqref{Eq:Boltzmann1stMom} that bears the physical meaning and not the definition of the stress tensor,  and (ii) our Eq.~\ref{Eq:StressTensor} remains the same for every form of $C$,  which as we will see later depends on the model of active matter.   As an example,  if $C=\partial_{\mu}f F_{\mu}$ with $F_{\mu}$ being proportional to a constant active propelled force,   a modified stress tensor could be defined as 
\bqn
P'_{\mu\nu} = P_{\mu\nu}  + F_{\nu} \left(\rho \bar{v}_{\mu}\right). 
\eqn
Finally,  the pressure is defined as the trace of the stress tensor divided by the dimension of the real-space
\bqn
P = \frac{1}{d} P_{\mu\mu},
\eqn
where there is a sum over the repeated indices.

The time evolution of the polarization vector of the particles is given by the first orientation moment of Eq.~\eqref{Eq:BoltzmannEq}
\bqn
\lb{Eq:Boltzmann_SMom}
\partial_t\left(\rho p_{\mu}\right)+ \partial_{\nu}\Gamma_{\mu\nu} 
-\int dv\, d\hat{u}\,d\omega\,  f\, \omega_{\nu}
= \int  dv\, d\hat{u}\,d\omega\,  \hat{u}_{\mu}C,\nb\\
\eqn
where the polarization vector is defined as
\bqn
\lb{Eq:PolarizationVec}
p_{\mu}\left(\vec{x},t\right)\equiv \frac{1}{\rho} \int  dv\, d\hat{u}\,d\omega\, f \hat{u}_{\mu},
\eqn
and we have defined 
\bqn
\Gamma_{\mu\nu}\left(\vec{x},t\right) \equiv  \int  dv\, d\hat{u}\,d\omega\, f \hat{u}_{\mu} v_{\nu}.
\eqn
A special but common model would be if the probability distribution of velocities and orientations are independent such that $f = f_v f_s$.  In this case,  $\Gamma_{\mu\nu} \propto p_{\mu}\bar{v}_{\nu}$.

Another order parameter of interest is the nematic tensor defined as 
\bqn
\lb{Eq:NematicTensor}
Q_{\mu\nu}\left(\vec{x},t\right)\equiv \frac{1}{\rho}\int  dv\, d\hat{u}\,d\omega\, f \left(\hat{u}_{\mu} \hat{u}_{\nu} -\frac{1}{d}\delta_{\mu\nu}\right),
\eqn
whose time evolution is given by the second orientation moment of Eq.~\eqref{Eq:BoltzmannEq}
\bqn
\lb{Eq:Boltzmann2ndSMom}
&&\partial_t\left(\rho Q_{\mu\nu} + \frac{\rho \delta_{\mu\nu}}{d}\right) + \partial_{\alpha} \int  dv\, d\hat{u}\,d\omega\, f \hat{u}_{\mu} \hat{u}_{\nu} v_{\alpha}\nb\\
&&-\int  dv\, d\hat{u}\,d\omega\, f \left(\omega_{\mu}\hat{u}_{\nu}+ \omega_{\nu}\hat{u}_{\mu}\right)
 =
\int  dv\, d\hat{u}\,d\omega\, \hat{u}_{\mu} \hat{u}_{\nu} C.\nb\\
\eqn
Again, under the special case that $f = f_v f_s$,  the integral in the second term on the left-hand side is proportional to $\bar{v}_{\alpha}\left(\rho Q_{\mu\nu} + \frac{\rho \delta_{\mu\nu}}{d}\right)$.

Since the one-body phase-space density is a probability function, we
use the Shannon definition of entropy as a measure of information in the system of particles through
\bqn
\lb{Eq:Entropy}
S\left(\vec{x},t\right) = - \int dv\, d\hat{u}\,d\omega\, f \ln f.
\eqn
It is interesting to note that when the Boltzmann H-theorem is valid,
the equation above is also equal to the thermodynamic entropy of the
system.  Additionally, when the interactions are strong, while the
above is a subset of the total thermodynamic entropy, it may still be an insightful measure of the system.

It is key to note that the goal of solving the hydrodynamic field equations~(\ref{Eq:Boltzmann0thMom},\ref{Eq:Boltzmann1stMom},\ref{Eq:Boltzmann_SMom},\ref{Eq:Boltzmann2ndSMom}) is to find $\rho\left(\vec{x},t\right)$, $\bar{v}_{\mu}\left(\vec{x},t\right)$, $p_{\mu}\left(\vec{x},t\right)$,  $Q_{\mu\nu}\left(\vec{x},t\right)$.  
The difficulty is that although the hydrodynamic equations above are
exact, they do not make a closed set of differential equations and
depend on higher order moments.  Therefore, if analytic solutions are
of interest, one has to make assumptions to break the hierarchy.  As can be seen from equations~(\ref{Eq:densityAndVelocity}, \ref{Eq:StressTensor},\ref{Eq:PolarizationVec},\ref{Eq:NematicTensor}),  the time and space dependence of all the order parameters above are known if $f$,  the one-body phase-space density, can be found.  
Here, we compute the left-hand side of Eq.~\eqref{Eq:BoltzmannEq} directly from data to arrive at $C$ without needing to break the hierarchy. 

\subsection{Analytic approaches}

Even though Eq.~\eqref{Eq:BoltzmannEq} is exact, it is not closed because $C$ is a functional of $f_2,  f_3,  \cdots$.  
Therefore,  analytic descriptions for $f$ is not possible unless we make simplifying assumptions.
A large subset of theoretical approaches to understanding active matter consists of proposing a form for $C$ based on a set of assumed symmetries and interactions.  In a wide class of models for active matter,  $C$ is given by the so-called Boltzmann equation such that 
\bqn 
\lb{Eq:C_Boltzmann}
C = I_{\text{dif}}[f] + I_{\text{col}}[f],
\eqn
where the detailed description of the two terms for active matter can be found in for example~\cite{PhysRevLett.109.268701,  Peshkov2014,  PhysRevE.74.022101, Bertin_2009}.  The Boltzmann equation have been used by numerous researchers to describe a wide range of systems of active matter, see for example \cite{PhysRevX.4.041030,Denk31623,PhysRevLett.120.258002}. 

Another prevalent class of models for active matter are considered to be explained by the Smoluchowski equation with its simplifying assumptions. 
For two dimensional self-propelled rods,  $C$ in Eq.~\eqref{Eq:BoltzmannEq} reads \cite{Baskaran_2010, PhysRevLett.101.268101} 
\bqn
C = D_R \partial^2_{\theta} f - \partial_{\theta}\left(f \tau \right) - \partial_{\mu} \left(f F_{\mu}\right),
\eqn
which is a slightly modified Smoluchowski equation, 
where $D_R$ is a noise coefficient,  $\theta$ is the angle of the orientation vector,  $\tau$ is the mean-field torque from other particles,  and $F$ is a mean-field force.  The original Smoluchowski equation of single-particle distribution function $f$, which again falls under the form of Eq.~\eqref{Eq:BoltzmannEq},  has been used to describe active cytoskeletal filaments \cite{PhysRevLett.96.258103}.  
Many other assumptions regarding the form of $C$ can be found in the literature.  For example,  in \cite{2020PhRvE.101b2602J},  the Gay-Berne potential is used to suggest a form for the time evolution of the one-body probability density $f$.  

One can also write the most general form for $C$ by constructing all possible scalars out of  $f$,  $v_{\mu}$,  $\hat{u}_{\mu}$,  $\partial_{\mu}$,  $\frac{\partial}{\partial_{v^{\mu}}}$,  $\frac{\partial}{\partial_{\hat{u}^{\mu}}}$,  etc.   The most general form of $C$ is the following:
\bqn
\lb{Eq:TheMostGeneralC}
C &=& a_1 \partial^2 f + a_2 \partial_{\mu} f v_{\mu} + a_3 \partial_{\mu} f \hat{u}_{\mu} + \cdots \nb\\
&+& \sum_{i=1}^{\infty} a_{(5,i)} (\sqrt{p_{\mu}p_{\mu}})^i 
+ \sum_{i=1}^{\infty} a_{(6,i)} (\sqrt{\hat{u}_{\mu}\hat{u}_{\mu}})^i \nb\\
&+& \sum_{i=1}^{\infty} a_{(7,i)} (\sqrt{\partial_{\mu}f\partial_{\mu}f})^i
+
\cdots,\nb\\
\eqn
where $a$ with any type of subscript stands for a passive or an active coefficient.  As can be seen,  we have an infinite number of models for active matter each corresponding with a $C$ equal to a subset of the terms above.  Nevertheless, for most of the commonly known models of active matter, only a few terms are non-zero.  
For example, $a_1$ in Eq.~\eqref{Eq:TheMostGeneralC} is similar to the viscosity in the Navier–Stokes equation,  and $a_3$ is similar to one of the active parameters in the dry flock model presented in Ref. \cite{Marchetti_2013},  where the moment of the right-hand side of Eq.~\eqref{Eq:BoltzmannEq} reads
\bqn
\int  dv\, d\hat{u}\,d\omega\, a_3 \partial_{\mu} f \hat{u}_{\mu} = a_3 \partial_{\mu}\left(\rho p_{\mu}\right). 
\eqn
In fact, this active-current term can be constructed within the Boltzmann framework of Eq.~\eqref{Eq:C_Boltzmann} as is shown in Ref. \cite{Bertin_2009}. 

It is also possible to impose constraints on the moments of Eq.~\eqref{Eq:BoltzmannEq} to arrive at a certain arrangement of non-zero coefficients in Eq.~\eqref{Eq:TheMostGeneralC}.  
One possible route is to study the terms analytically by categorizing them according to some group properties. In that way,  we will end up with a study similar to the one presented in ~\cite{Toner_Tu_PRL_1998,Toner_Tu_PRE_1998}.   In this paper, we follow a different approach to be discussed next.

\subsection{Data-driven approach}
\lb{Sec:EstPhaseSpace}
As we mentioned above, $C$ accounts for all of the higher-body probability functions such that Eq.~\eqref{Eq:BoltzmannEq} is exact.  However,  simplifying approximations are needed to find an analytic solution for $f$ because the aforementioned equation is not closed. 
An alternative to making simplifying assumptions,  is to directly estimate $f$ from data  and use the exact form of Eq.~\eqref{Eq:BoltzmannEq} to find $C$ as well as the order parameters such as $\rho\left(\vec{x},t\right)$, $\bar{v}_{\mu}\left(\vec{x},t\right)$, $p_{\mu}\left(\vec{x},t\right)$,  and $Q_{\mu\nu}\left(\vec{x},t\right)$.

\begin{figure}
\centering
\includegraphics[width=0.8\columnwidth]{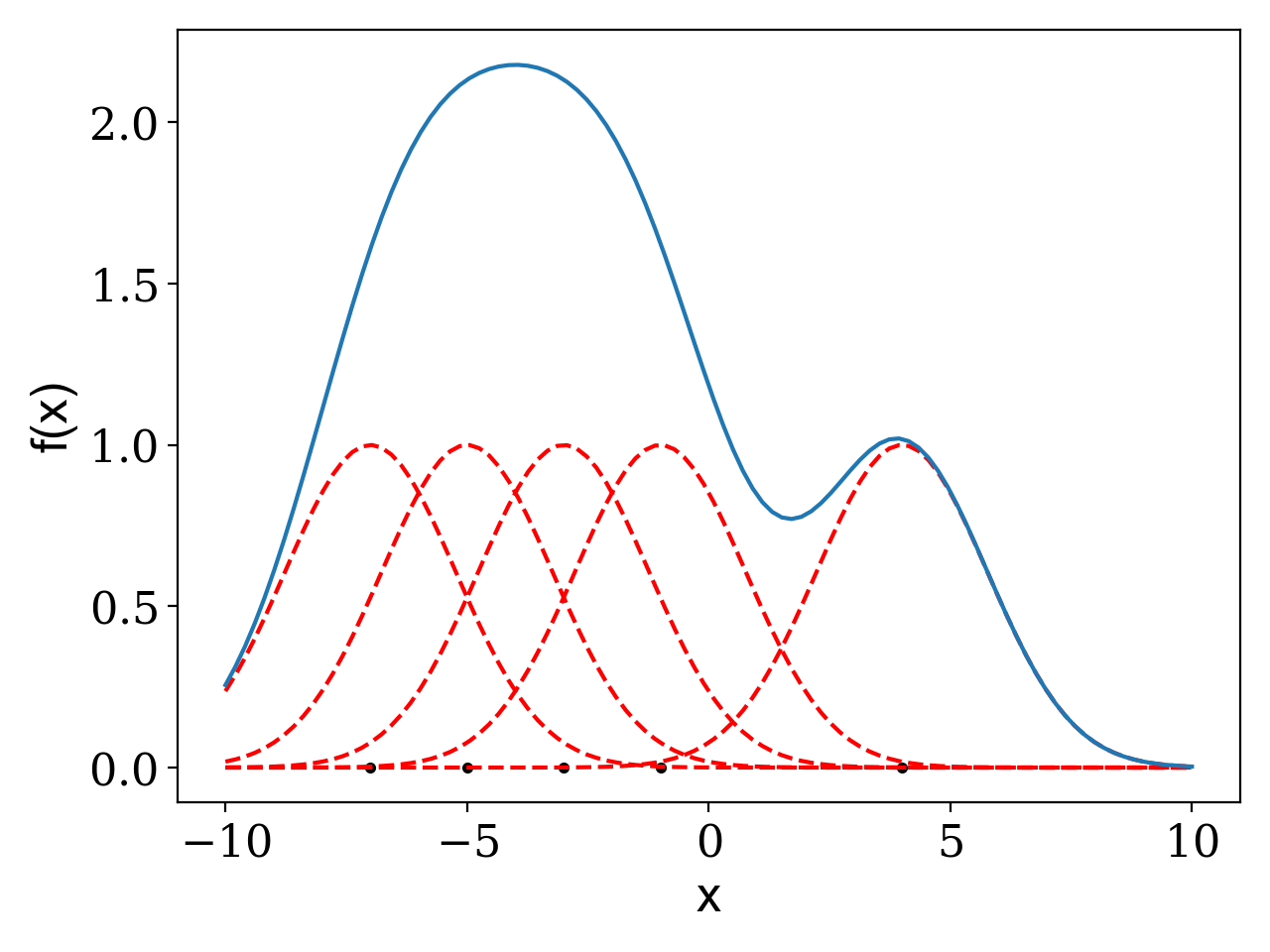}
\includegraphics[width=\columnwidth]{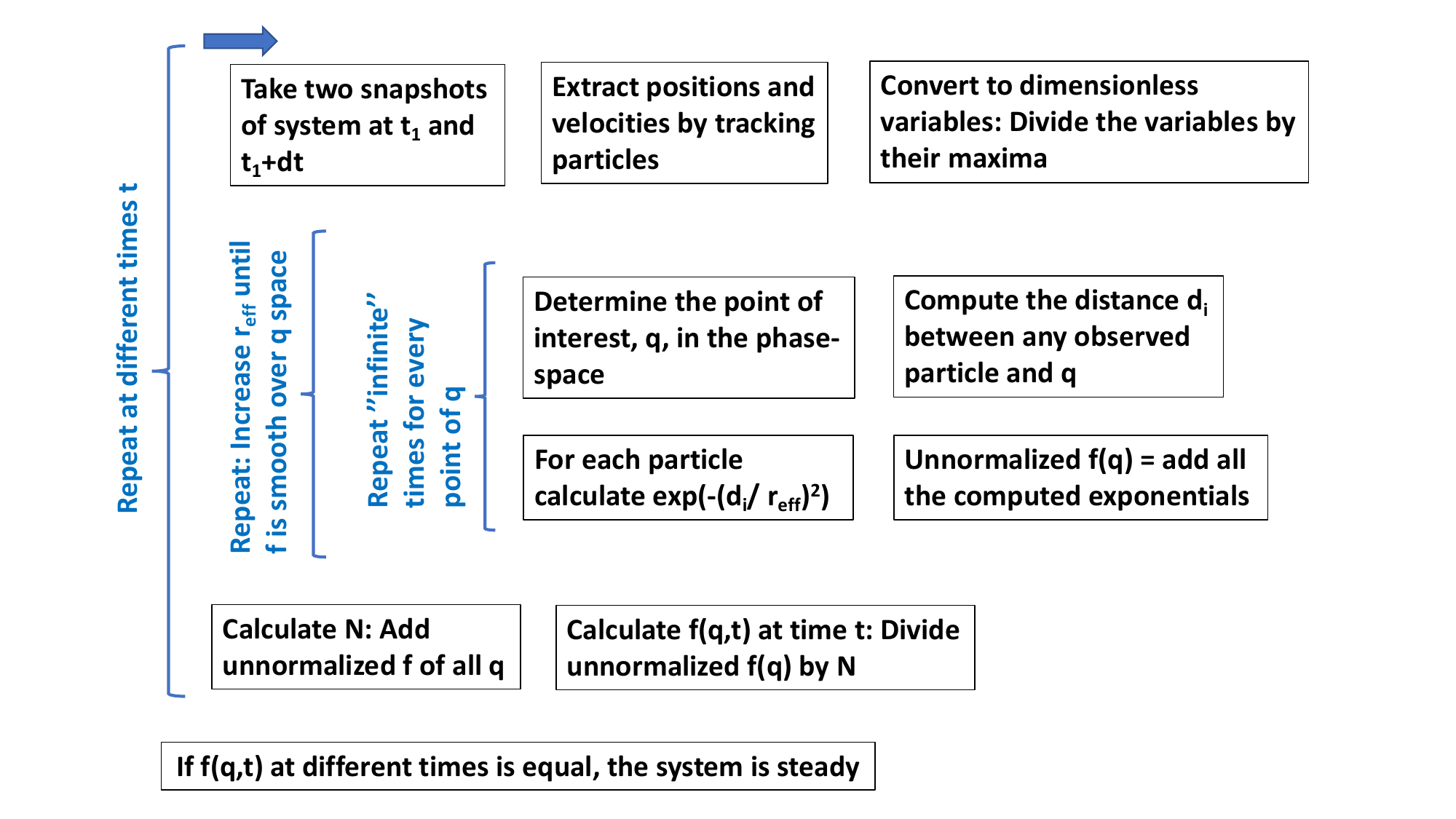}
\caption{(Top) panel shows the density estimator in a one dimensional
  phase-space.  There are only five data points shown with black dots.
  Their contribution to the density function is denoted by the red
  dashed lines.  The sum of all the contributions is represented by the
  solid blue line and serves as the estimation of the unnormalized probability function.   
(Bottom) panel is a pictorial demonstration of the algorithm of this data-driven method. \lb{Fig:AlgChart}}
\end{figure}
Biological systems are often made of many types of particles and we should write one Eq.~\eqref{Eq:BoltzmannEq} for every particle type.  In this case, $C$ not only accounts for higher-body distribution functions but also for the interactions between different types of particles.  In the following,  we assume that only one type of particle is of interest.  We also assume that an experiment has collected a data sample
containing the information of positions and orientations of the particle type of interest over time.  

The starting point would be to note that the one-body phase-space density $f(\q)$ is equal to the probability of finding a particle at point $\q$ in phase-space. 
Perhaps the most straightforward approach to estimating $f(\q)$ from a
data sample is to discretize the $\q$ space and count the number of
particles that fall into each bin. The problem with this approach is
that if the number of dimensions of $\q$ is high, most of the bins remain
empty. 
For example,  a typical dataset contains only a few
hundred to a few ten-thousand particles. 
 On the other hand,  
even for a poor and blurry discretization of 10
bins per dimension, 
and for a typical six-dimensional phase-space,  
we end up with one million bins. 
Another problem with this approach is that
the estimation often depends on the choice of the locations of the bin edges. 

Although the simple approach above is not necessarily practical for estimating the phase-space density, its careful interpretation can guide us in devising a better estimator. An interpretation of the binning method is that all of the observed particles contribute to a given bin with binary weights. In other words, a particle's contribution to a given bin is zero or one depending on whether the phase-space distance of that particle to the center of the bin is greater or smaller than the bin size. The estimation of the phase-space density then reads
\bqn
\lb{Eq:PhasSpacDensEst}
f(\q) = \Ni \sum_{i} w_i(\q),
\eqn 
where $i$ runs over particles, $\Ni$ is a normalization factor, and $w_i(\q)$ is the contribution of the $i$th particle to the probability at $\q$. 

It is already obvious that the binary form of $w_i(\q)$ is behind the problems of the binning method,  and $w_i(\q)$ should be a continuous function such that all of the observed particles have a non-zero contribution to a given location of phase-space.  
In statistics,  Eq.~\eqref{Eq:PhasSpacDensEst} with a continuous weight factor is known as kernel density estimator \cite{10.1214/aoms/1177728190}. 
Also, the weights of particles close to a given phase-space location $\q$ should be more significant than the weight of particles that are far from it. Therefore, we use the following Gaussian form for the weights in Eq.~\ref{Eq:PhasSpacDensEst} 
\bqn
\lb{Eq:W_i}
w_i(\q) = \exp\left(-\left(\frac{|\q -\q_i|}{\rf}\right)^2\right),
\eqn
where $\rf$ is a free parameter.  
The optimized value for $\rf$ would be the smallest number that still returns a smooth estimation of phase-space.  
Hence, by observing the position of particles in the phase-space $\q_i$,  the weight $w_i(\q)$, and subsequently the density $f(\q)$ can be estimated. 
Finally, phase-space density reads 
\bqn
f(\q,t) = \Ni \sum_{i} \exp\left(-\left(\frac{|\q -\q_i(t)|}{\rf}\right)^2\right). 
\eqn
A pictorial description of the algorithm for our estimator of the phase-space density can be found in Fig.~\ref{Fig:AlgChart}.   

By inserting this continuous one-body phase-space density function into the equations~(\ref{Eq:densityAndVelocity} ,  \ref{Eq:StressTensor},  \ref{Eq:PolarizationVec},  \ref{Eq:NematicTensor},  \ref{Eq:Entropy}),  we can estimate the number density,  the bulk velocity,  the stress tensor,  the polarization vector,  the nematic tensor,  and the entropy as functions of time,  and as the solutions of the hydrodynamic equations in section~\ref{Sec:Hydro} without needing to solve the differential equations.

Having estimated the phase-space density at different times $t$,  and assuming that the external forces are either known or absent,  all of the terms on the left-hand side of Eq.~\ref{Eq:BoltzmannEq} are known.  Therefore,  we can find an estimation for $C$,  which can be subsequently compared with Eq.~\eqref{Eq:TheMostGeneralC} to find its significant terms.

\subsection{Statistical field theory of active matter}
\lb{Sec:StatFieldTheory}  
A field theoretic description of a system is potentially powerful due to its capability in explaining various experiments that are seemingly different.  It is this one-to-many relationship that highlights the importance of field theory. 
When an active matter is close to its steady-state, an effective statistical field theory of its order parameters can be constructed. 
An objective of this paper is to learn this effective field theory by observing one of the experiments that it can explain.  

To start the construction,  we note that,  on the one hand,  the correlations between the order parameters are determined by a probability functional,  and if these correlations are known,  one should be able to recover the probability functional through a reverse engineering.  On the other hand,  in Sec.~\ref{Sec:Theory},  we have devised a method to learn the space and time dependence of the order parameters,  and consequently their correlations,  from data.  
More specifically,  by definition,  the effective partition function is the sum of the probabilities of possible configurations
\bqn
\lb{Eq:TotalPartition}
&&Z[J] = \int \D\varphi\, \exp\bigg(-S_{\text{total}}[\varphi]-\int d^3x \varphi J  \bigg),
\eqn
where $\varphi$ represents the order parameters,  and $J$ is an auxiliary field. 
On the one hand,  the correlations among the order parameters are given by 
\bqn
\lb{Eq:CorrFunction}
&&\langle \varphi(\vec{x}_1)\cdots \varphi(\vec{x}_k) \rangle 
=\frac{\delta^k Z}{\delta J(\vec{x}_1)\cdots \delta J(\vec{x}_k)}\Big|_{J=0}.
\eqn
On the other hand,  the same correlations are known through the data-driven method of Sec.  \ref{Sec:Theory}.  The objective is to construct the unknown and analytic expression on the right-hand side from the left-hand side,  which is known numerically.

\lb{Sec:Greens}
In the following,  we pursue the construction of the effective field theory by first expanding the right-hand side of Eq.~\eqref{Eq:CorrFunction} using the perturbation theory,  which is valid given the assumed stationary state of the system.  Each term in the expansion would be an unknown variable to be determined.  Therefore,  we need to observe as many correlation functions,  for k = 2,  3,  $\cdots$,  as the number of unknown variables.  By solving the system of equations,  the leading terms in the expansion of the partition function would be recovered from data.  It should be noted that,  usually,  a few first terms of the expansion are of interest, and the higher-order terms can be neglected for the sake of practicality and depending on how much data is available.

To practically use the right-hand side of Eq.~\eqref{Eq:CorrFunction},  we need to calculate the path integration in Eq.~\eqref{Eq:TotalPartition},  and write $Z[J]$ explicitly in terms of the leading perturbations in $S_{\text{total}}[\varphi]$.
Since the system is close to the equilibrium and the fluctuations around the saddle point are small, 
the exponential function in Eq.~\eqref{Eq:TotalPartition}, equivalent to the probability of configurations, takes the following form
\bqn
\lb{Eq:ProbabilityCloseToEquil}
&&{\cal{P}} = \nb\\
&&\exp\Bigg(-\frac{1}{2}\int d^3x_1 d^3x_2 
\frac{\delta^2 S_{\text{total}}}{\delta \varphi(x_1)\delta \varphi(x_2)}\Big|_{\varphi_{_{\text{SP}}}}
\varphi(\vec{x}_1)\varphi(\vec{x}_2)\nb\\
 &&+ V(\varphi) - \int d^3x\varphi(\vec{x}) J(\vec{x})   \Bigg),\nb\\
\eqn
where $V(\varphi)$ refers to the higher order terms, and we have assumed that at the saddle point $\frac{\delta Z}{\delta \varphi}\Big|_{\varphi_{_{\text{SP}}}}=0$. 

To recover the leading term,  we introduce subscript ``0'' to refer to $V(\varphi)=0$, and define $\varphi_0(x)$ to be the solution to the corresponding field equation  
\bqn
\int d^3x' 
\frac{\delta^2 S_{\text{total}}}{\delta \varphi(x')\delta \varphi(x)}\Big|_{\varphi_{_{\text{SP}}}}
\varphi_{_0}(\vec{x}') = - J(\vec{x}),
\eqn
which can be used to calculate the non-interacting partition function
\bqn
\lb{Eq:Z_0J}
Z_0[J] = \exp\left(-\frac{1}{2}\int d^3x_1 d^3x_2 J(\vec{x}_1)J(\vec{x}_2) \Delta(\vec{x}_1-\vec{x}_2)\right).\nb\\
\eqn
Here, we have dropped a normalization factor, and $\Delta\left(\vec{x}_1-\vec{x}_2\right)$ is the Greens' function defined as 
\bqn
\lb{Eq:GreenFunctionEquation}
\int d^3x' 
\frac{\delta^2 S_{\text{total}}}{\delta \varphi(x')\delta \varphi(x_1)}\Big|_{\varphi_{_{\text{SP}}}}
\Delta(\vec{x}'-\vec{x}_2) =\delta(\vec{x}_1-\vec{x}_2).
\eqn
Finally,  we achieve the desired expansion of the full partition function through a straightforward calculation
\bqn
Z[J] &=& \exp\left( V\left(\frac{\delta}{\delta J}\right)\right)Z_0[J]\nb\\
&=& \left(1 +  V\left(\frac{\delta}{\delta J}\right) + \cdots \right) Z_0[J].
\eqn

The algorithm for reconstructing the effective action is as follows.
First, we use the data-driven method,  developed in
Sec.~\ref{Sec:Theory},  to find the phase-space density at different
moments and determine whether or not $f$ evolves with time within some
tolerance.  If not, we collect many snapshots of the system and estimate the phase-space density for each of them.  The phase-space densities will be used to learn the order parameters as functions of the real-space position in the corresponding snapshots.  
The snapshots act as the members of the ensemble of the system.  Correlations between the order parameters of different locations,  such as $\langle \varphi(\vec{x}_1)\cdots \varphi(\vec{x}_k) \rangle$,   are equal to the average of the multiplications of order parameters,  i.e.  $\varphi(\vec{x}_1)\cdots \varphi(\vec{x}_k)$,  over the snapshots.

\begin{figure*}
\centering
\subfigure[]{\includegraphics[width=\columnwidth]{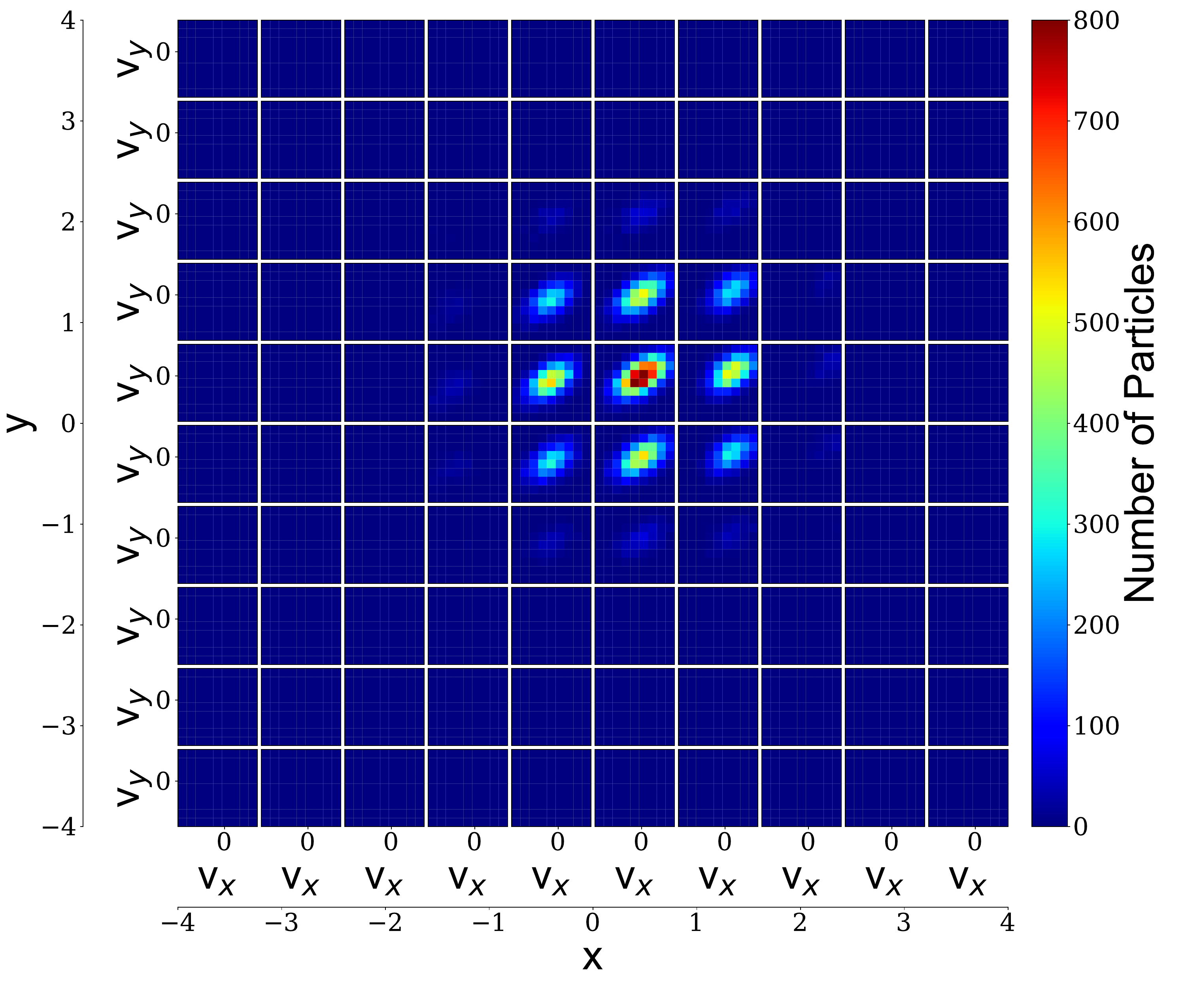}}
\subfigure[]{\includegraphics[width=\columnwidth]{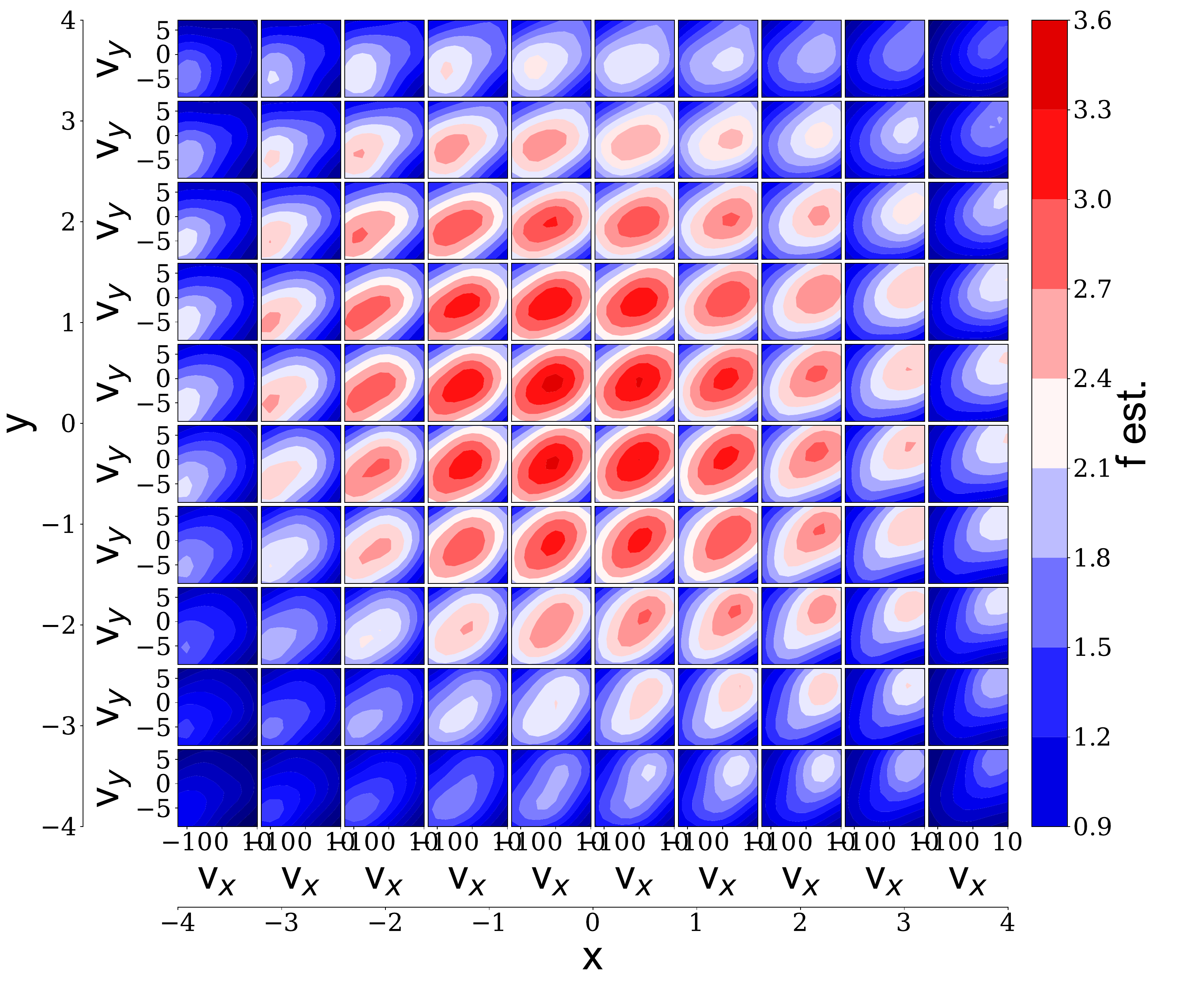}}
\subfigure[]{\includegraphics[width=0.9\columnwidth]{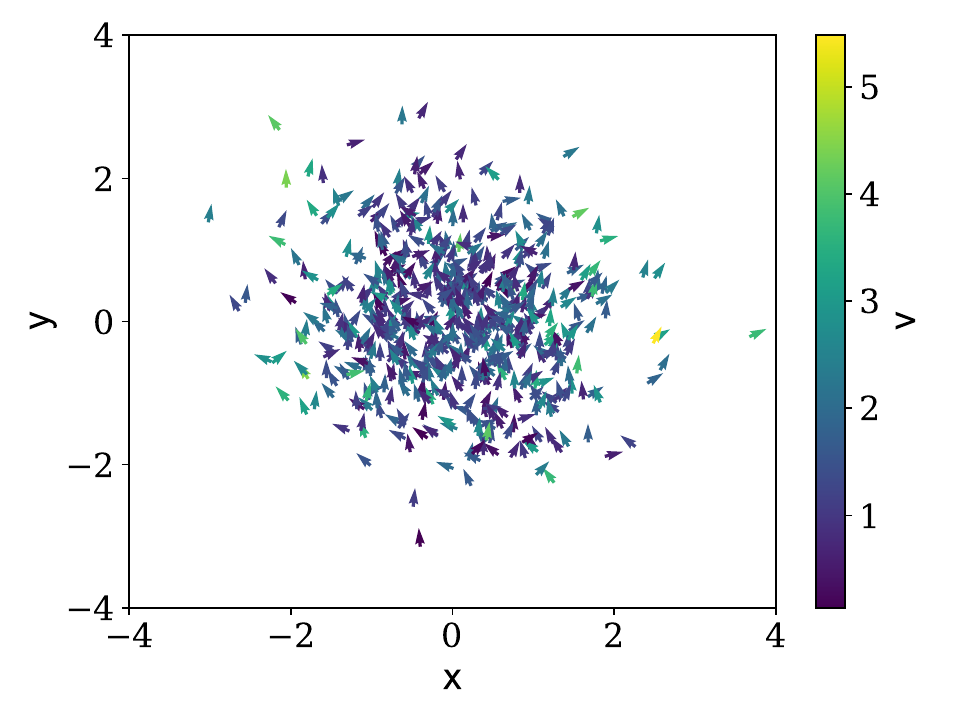}}
\subfigure[]{\includegraphics[width=0.9\columnwidth]{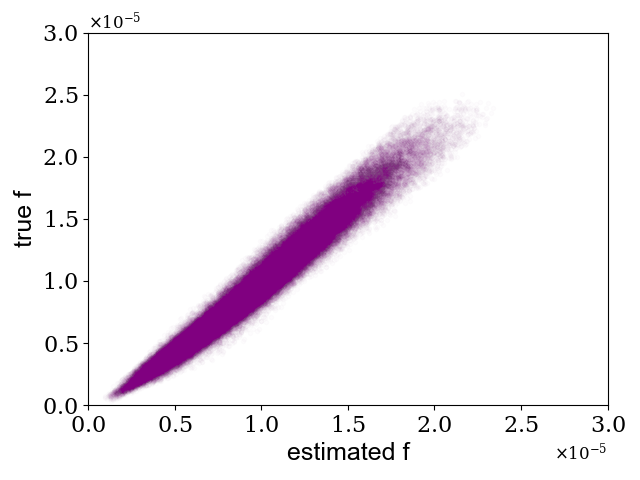}}
\caption{Simulation test case I: 
Simulation of a five-dimensional dataset including two real-space, two
velocity space, and one angle of direction for 1000 particles. We have assumed the true phase-space density and used it to generate the data. The data are passed to our method to estimate the phase-space density. This plot shows that our estimation of the phase-space density is quite accurate. The plot also shows that a simple histogram of the positions of the 1000 particles cannot appropriately represent the underlying phase-space density.
(a) Four dimensional presentation of the position of the simulated particles in phase-space in one single frame.
At each $x-y$ position, there exists a subplot of $v_x-v_y$ space that
shows the color value of the numbers. Due to the low statistics in the
simulations, which is typical in the real experiments, most of the
regions of the plot are empty.  \lb{Fig:Sim4DplotsBacteria}
(b) Four dimensional presentation of our estimation of the phase-space number density $f(\q)$.
At each $x-y$ position, there exists a subplot of $v_x-v_y$ space that
shows the color value of the densities. This plot shows that the true
value of the probability of finding particles at the empty regions of
the histogram in the previous panel is not zero. This difference
between observed particles and the true probability distribution is
addressed by our method \lb{Fig:Sim4DplotsBacteria}
(c) Snapshot of the simulated data. The color bar shows the velocity
of particles. The video clip of this dataset can be seen at \href{https://www.dropbox.com/s/lpslxazn30cszi1/SimData.mp4?dl=0}{here}.  \lb{Fig:dataset1}
(d) Scatter plot of the five-dimensional phase-space density. The y-axis shows the true values and the x-axis shows the corresponding estimated values for the choice of $\rf=0.5$. 
\lb{Fig:f_comparisons1}
} 
\end{figure*}
\begin{figure*}
\centering
\subfigure[]{\includegraphics[width=\columnwidth]{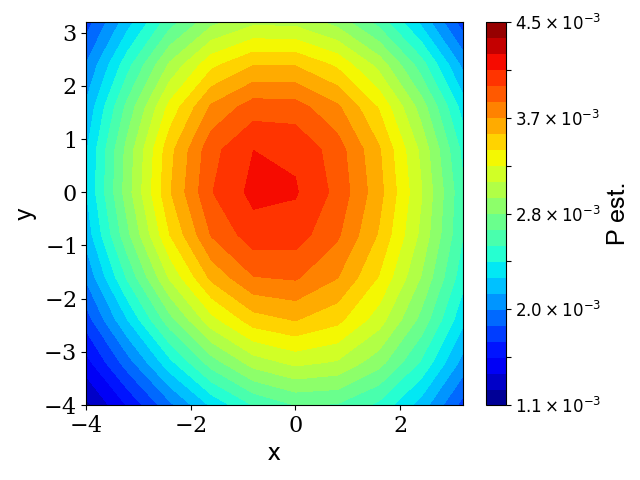}}
\subfigure[]{\includegraphics[width=\columnwidth]{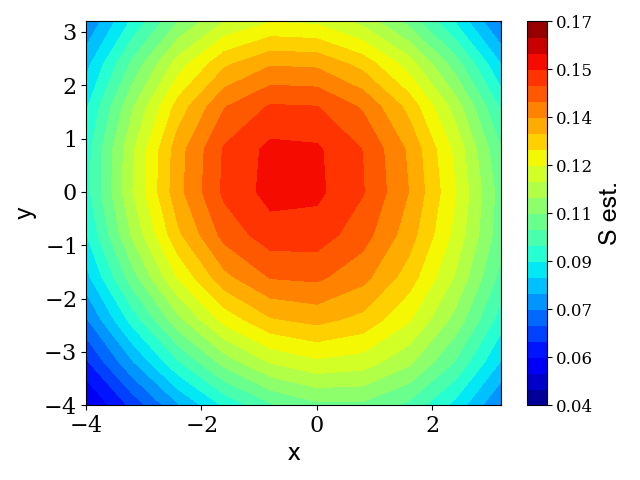}}
\caption{
(a) The estimated pressure for simulation test case I. 
(b) The estimated entropy for simulation test case I. \lb{Fig:PressureEntropy}
} 
\end{figure*}
\section{Simulation test cases}
\lb{Sec:Simulations}
To test the data-driven approach outlined in Sec.~\ref{Sec:Theory}, we apply it to datasets whose properties are
known. We consider the following three test cases. 

\subsection{Simulation test case I: Multivariate Gaussian distributions}
In this subsection,  we assume the following five-dimensional phase-space distribution function
\bqn
\lb{Eq:DistSim1}
f(\q) = \Ni \exp\left(-\frac{1}{2}\q\cdot \Sigma^{-1}\cdot \q\right),
\eqn
where $\q=(x, y, v_x,v_y,\cos\theta)$ with $\theta$ being the angle of
orientation for rod-like particles, and 
\bqn
\Sigma^{-1} = 
\begin{bmatrix}
  1.9 & 0.18 & -0.66 & -0.23 & -0.43\\
  0.18 & 1.18 & 0. & 0. & -0.43\\
  -0.66 & 0. & 1. & -0.45 & 0.\\
  -0.23 & 0. & -0.45 & 1.2 & 0.\\
  -0.43 & -0.43 & 0. & 0. & 1.\\
\end{bmatrix}.
\eqn
Note that the Maxwell-Boltzmann distribution is a very special case of the distribution above when every component of $\Sigma^{-1}$ is zero except $\Sigma^{-1}_{33}=\Sigma^{-1}_{44}\neq 0$. It should also be noted that many interactions have been assumed between positions, velocities, and polarization, and so the system is far from a non-interactive one.

We use the ``numpy.random.multivariate\_normal'' package in python to draw 15 snapshots, each with 1000 particles.   
A histogram of the four-dimensional phase-space position of observed particles, which is the naive binning method discussed above, is shown in Fig.~\ref{Fig:dataset1}(a). The estimated phase-space density is shown in Fig.~\ref{Fig:dataset1}(b). The deficiency of the simple binning approach can be seen by comparing  Fig.~\ref{Fig:dataset1}(a) and Fig.~\ref{Fig:dataset1}(b).  
One of the snapshots can be seen in Fig.~\ref{Fig:dataset1}(c).
The true and the estimated, using $\rf=0.5$, phase-space densities at every point of the phase-space are compared. The result can be seen in Fig.~\ref{Fig:dataset1}(d). This panel shows that unlike the histogram approach, the method can accurately estimate the true distribution function.

We study the effects of tuning $\rf$ by estimating the phase-space density with different values of $\rf$ for each of the 15 samples and comparing their average as our final estimation with Eq.~\ref{Eq:DistSim1}. Fig.~\ref{Fig:SSS_dEff} shows the error in our estimation of the phase-space density in terms of $\rf$. As can be seen from the figure, $\rf\simeq 0.5$ returns the most accurate estimation of the phase-space density. 
On the other hand, Fig.~\ref{Fig:SSS_NSample} show that the error of estimation can be reduced by increasing the number of samples, snapshots of the system.

Having estimated the phase-space density using 15 samples with $\rf=0.5$, we are ready to compute the order parameters  of interest using Eqs.~(\ref{Eq:densityAndVelocity}, \ref{Eq:StressTensor},\ref{Eq:PolarizationVec},\ref{Eq:NematicTensor}).
For instance,  our estimations of the pressure and the entropy of the simulated data are shown in Fig.~\ref{Fig:PressureEntropy}
Also,  assuming that no significant external force exist in the system, we can estimate $C$ at every location of phase-space by inserting the estimated phase-space density into the left-hand side of Eq.~\eqref{Eq:BoltzmannEq} and find the significant terms in  Eq.~\eqref{Eq:TheMostGeneralC}.

\begin{figure}[thb]
\centering
\subfigure[]{
\includegraphics[width=0.48\columnwidth]{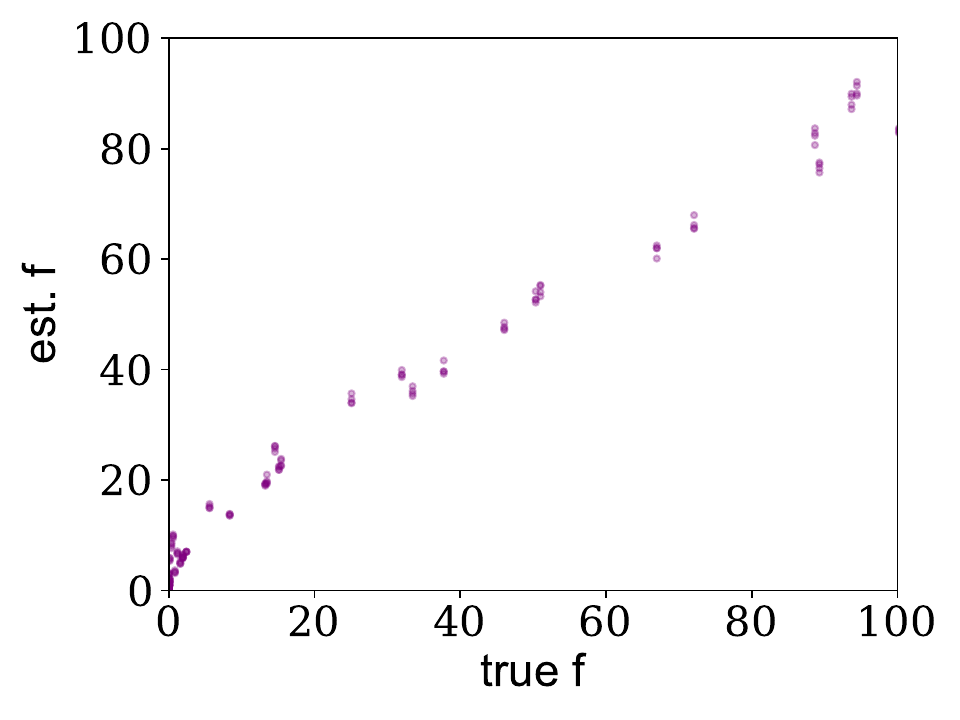}}
\subfigure[]{
\includegraphics[width=0.48\columnwidth]{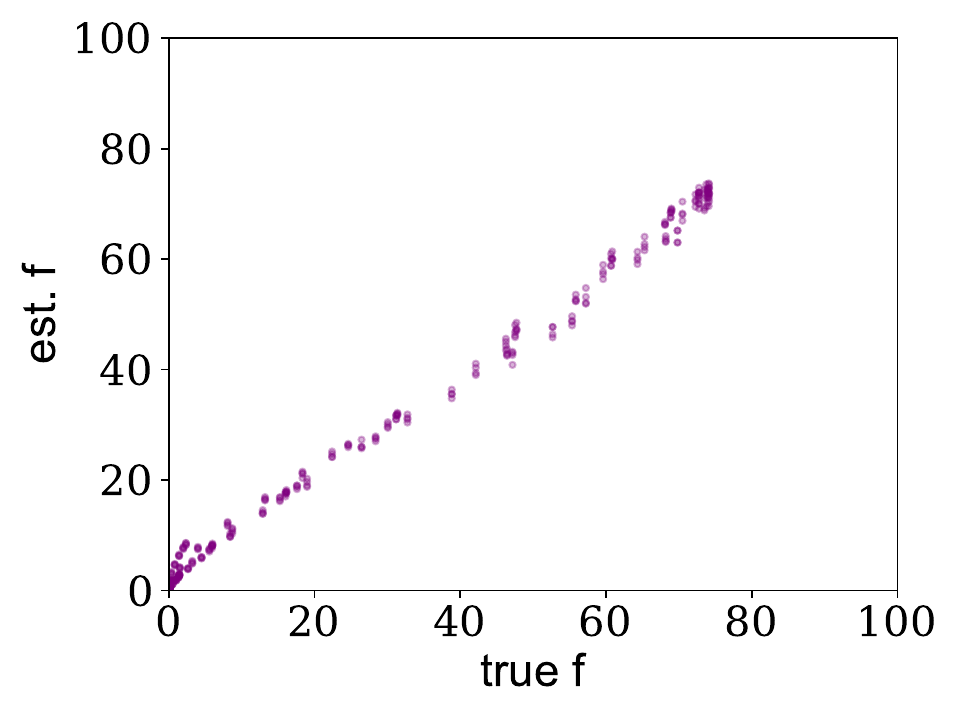}}
\subfigure[]{
\includegraphics[width=0.48\columnwidth]{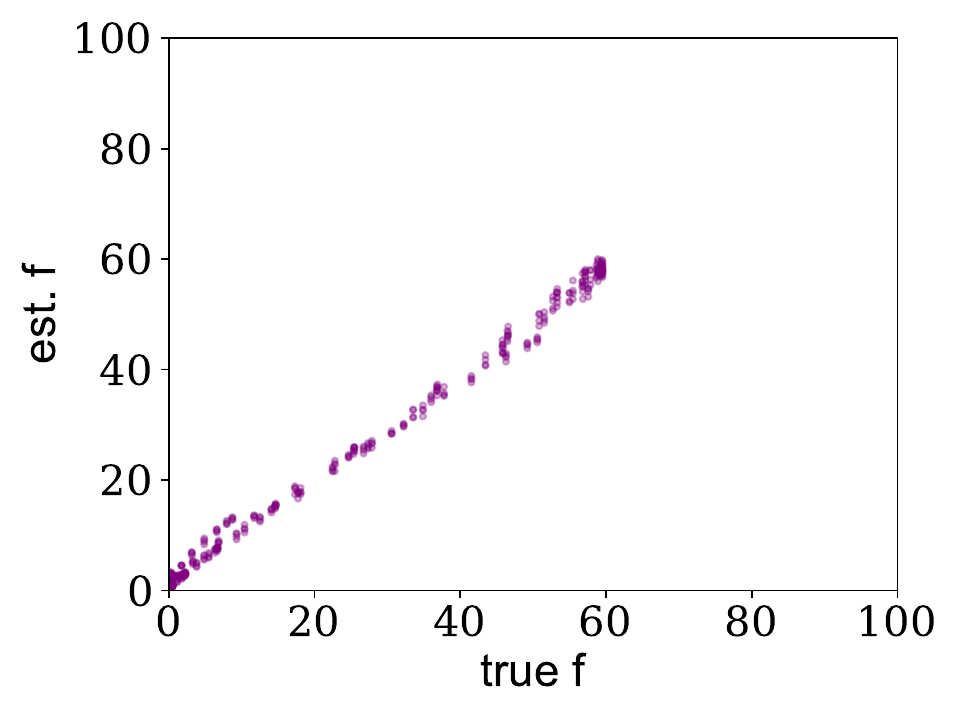}}
\subfigure[]{
\includegraphics[width=0.48\columnwidth]{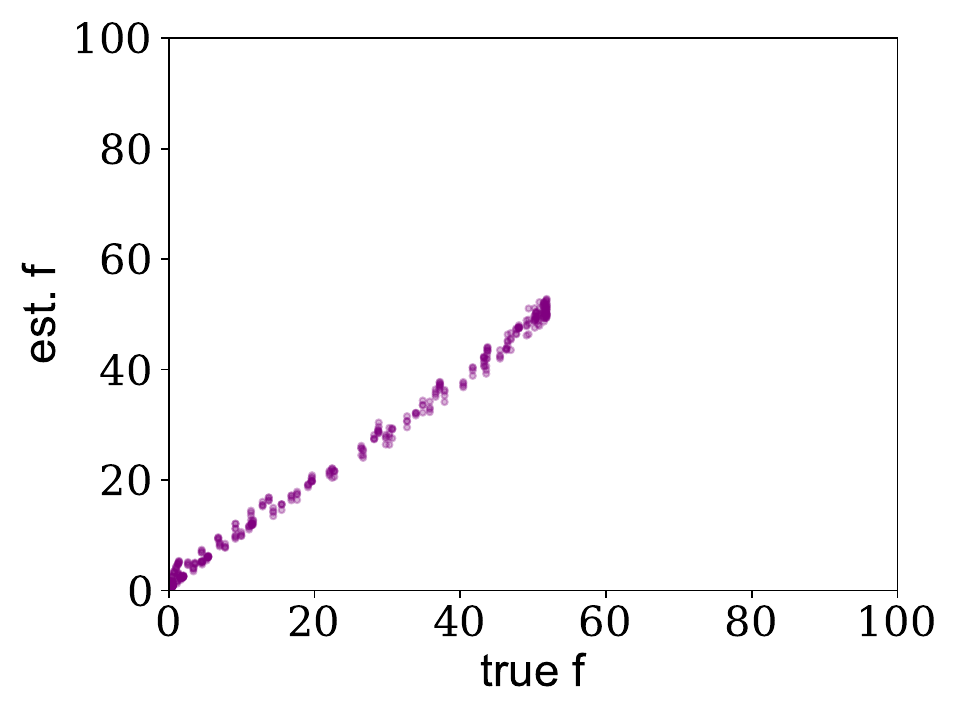}}
\caption{Simulation test case II: Comparison of the data-driven estimation of the phase-space
  density with the true one in a non-linear, non-steady-state system.  The true distribution is $f = {\cal{N}}\exp(-v^4/T)$ with the parameter $T=t^{\frac{3}{7}}$ changing with time such that $\frac{\partial f}{\partial t} \neq 0$.  We let the system evolve with time and take snapshots at four time points to make the estimation.  (a) $T = 0.3$,  (b) $T=1.5$, (c) $T=2.6$, (d) $T=3.8$.\lb{Fig:NonLinear}}
\end{figure}
\subsection{Simulation test case II: Non-linear, non-steady-state scenario}
To test our method in a non-linear and non-steady system,  we simulate a system of particles on a surface whose phase-space density is 
\bqn
\lb{Eq:fNonLinearSim} 
f(t,v_x, v_y) = {\cal{N}}\exp(-\frac{v^4}{t^{\frac{3}{7}}}).
\eqn

Although the distribution function has a non-conventional form, we are still able to draw samples using python's random package with the following trick.  First,  we use the ``numpy.random.uniform'' to generate one million data point in $v_x$ space.  We scale the velocity space by dividing $v_x$ by the length of the container of the particles such that the velocities are confined within an interval of (-4,4).  We repeat the same procedure to generate uniform points in the $v_y$ dimension.  By zipping the two arrays,  we have one million uniformly generated data point in the two-dimensional phase-space.  The number of uniformly generated data point is intentionally high to cover almost the entire phase-space.  Next,  we use Eq.~\eqref{Eq:fNonLinearSim} to compute the probability of each of the data point.  Finally,  we use ``random.choices'' package in python to draw samples of 10000 particles out of the one-million population based on their computed probability.  

To increase the accuracy of the estimation,  we take ten snapshots of the system at each time, corresponding to having ten replicates,  and repeat the procedure at four different times.  We insert the snapshots as an input to our data-driven method and estimate the phase-space density.  The average of the ten estimated phase-space densities at each time is compared with the true function above.  Figure~\ref{Fig:NonLinear} shows that our method accurately estimates such non-linear, and out of equilibrium systems.  We choose $\rf=0.05$ at all the time points to get the best optimized result.  

Since our estimation of $f(t,v_x,v_y)$ is closely similar to the true density at all of the time points, our estimation of any of the order parameters of the system will be accurate.  For example, the pressure of the system reads
\bqn
P(t) =  \frac{1}{2} \int \, d^2v f v^2 = \frac{{\cal{N}} \pi}{4}t^{\frac{3}{4}},
\eqn
which is clearly out of equilibrium and non-linear and solely depends on how well we can estimate $f(t,v_x,v_y)$.

\begin{figure*}[thb]
\centering
\subfigure[]{
\includegraphics[width=\columnwidth]{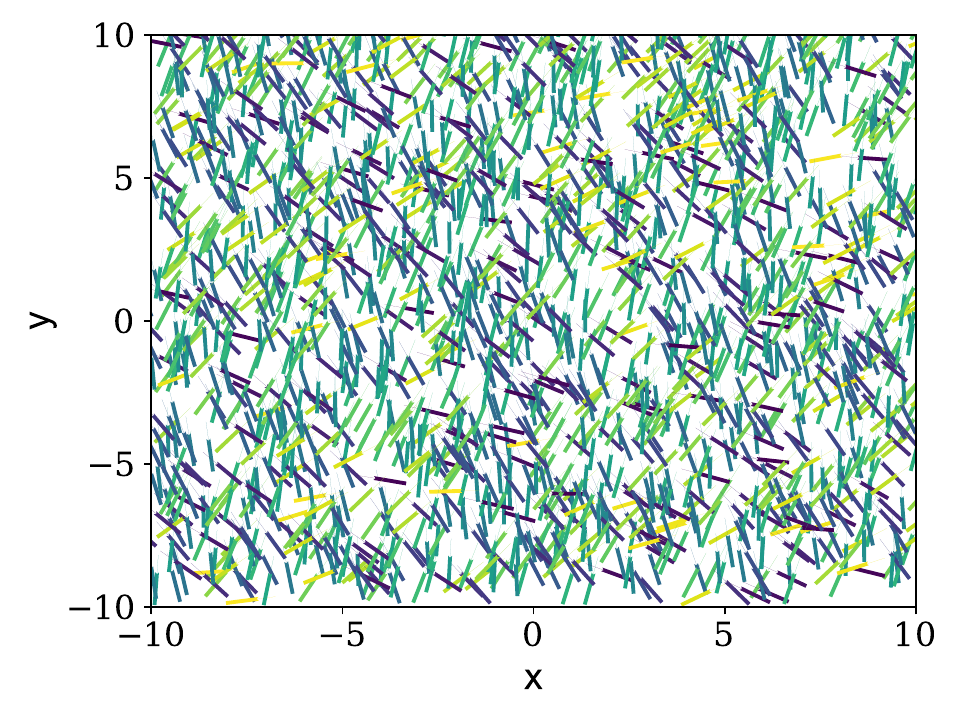}}
\subfigure[]{
\includegraphics[width=\columnwidth]{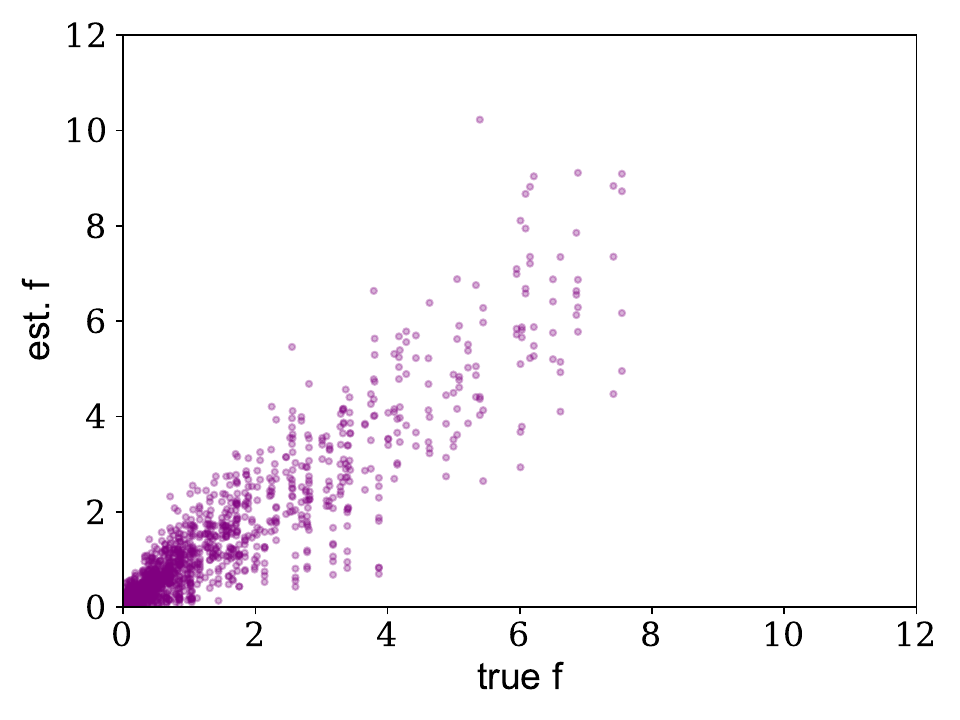}}
\caption{Simulation test case II: (a) One of the ten snapshots of the self-propelled rods
  system.  The parameters are chosen such that $\rho_0 > \rho_N$ with
  the director unit vector $\hat{n}$ in the y-direction such that
  there are more particles in the y-direction than those in the
  x-direction in accordance with the analytic solution in Eq.~\ref{Eq:analytic_f_self_prop_rod}. 
(b) The estimated versus the true phase-space density at uniformly
distributed positions of the 6-dimensional phase-space.   The
comparison shows fair agreement between the two.  Therefore, our
method's estimations for the additional order parameters are fairly
accurate too because these parameters are simply moments of the phase-space density $f$ as defined in the theory section above and also in \cite{PhysRevE.77.011920, PhysRevLett.101.268101}. 
\lb{Fig:RodSim}}
\end{figure*}
\subsection{Simulation test case III: Self-propelled rods}
We now evaluate the performance of our method by applying
it to a system whose analytic hydrodynamic solution is known, namely, self-propelled hard rods on a substrate in two dimensions.  Both the microscopic and continuum description of the system can be found in Refs.~
\cite{PhysRevE.77.011920, PhysRevLett.101.268101}, where the order
parameters such as the number density, the polarization vector, and
the nematic tensor are defined in terms of the one-particle phase-space density that satisfies a modified Smoluchowski equation of the general form of Eq.~\eqref{Eq:BoltzmannEq} where $q_{\mu}=\left(\vec{r},\theta, \vec{v},\omega \right)$ refers to the position vector in the two-dimensional configuration space, the angle that defines the orientation of the rods, the velocity of the rods, and their angular velocities.  
As discussed in the two references above, if the friction due to the
substrate is large and the interactions are of the excluded volume
type, and after a few more simplifying assumptions, the following would be a stationary solution
\bqn
\lb{Eq:analytic_f_self_prop_rod}
&&f\left(\vec{r},\theta, \vec{v},\omega \right) \simeq f_x(\vec{r},\theta) \times\nb\\
&&\exp\left(   
-\frac{1}{2 k T} \left( 
(\vec{v}-v_0 \hat{u})^2 + I \omega^2
	\right)
\right),
\eqn
where $v_0$ is the self-propulsion along the direction of the rod, and
$I =l^2/12$ with $l$ being the length of the rods. Moreover, 
$f_x$, the real-space probability distribution, takes the following form
\bqn
f_x = 
\begin{cases}
\frac{\rho_0}{2\pi},  & \rho_0 < \rho_N,\nb\\
\frac{\rho_0}{2\pi}\left( 1+ 4 Q_{\alpha\beta} \left(\hat{u}_{\alpha}\hat{u}_{\beta} - \frac{1}{2} \delta_{\alpha\beta}\right)\right), &
 \rho_0 > \rho_N,
\end{cases}
\eqn
where $\rho_0$ is a constant,  $\rho_N=\frac{3\pi}{2 l^2}$ is the Onsager transition density,  and 
$Q_{\alpha\beta}=S \left( \hat{n}_{\alpha} \hat{n}_{\beta} -\frac{1}{2}\delta_{\alpha\beta}\right)$,  with 
$S$ and $\hat{n}$ being constant scalar and unit vector.  For more detail, we refer to the two references above. 

Before we proceed, it should be clarified that although many
dynamical microscopic simulations exist in the literature, their
corresponding phase-space density is not necessarily known.  On the
other hand, to evaluate our method, we need to compare our estimated
phase-space density with the true one.  Fortunately,  this true
phase-space density of the system of self-propelled rods has been
analytically derived in Refs.~\cite{PhysRevE.77.011920,
  PhysRevLett.101.268101} starting from the underlying microscopic
system.   Therefore, we test our method in the same manner as described in test case II. 

In the following, we simulate this system by choosing the length of the rods as $l=1$,  the temperature as $kT=2$,  the active parameter as $v_0=2$,  and without loss of generality $\hat{n}=(0,1)$.  
Also,  we choose $\rho_0=5$ to satisfy $\rho_0>\rho_N$ since this is the more complex scenario whose estimation is harder. 
We take 10 snapshots of the system to serve as the input data into our data-driven method.  One snapshot of the system can be found in Fig.~\ref{Fig:RodSim}(a).

Next, we pass the snapshots to our data-driven method to estimate the
single-particle phase-space density.  We find that $\rf=0.1$ gives the
best optimized estimation.  We compare our estimation with the
analytic solution in Eq.~\eqref{Eq:analytic_f_self_prop_rod} and
validate the method in such active systems.
Figure~\ref{Fig:RodSim}(b) shows that since our method returns a fair
estimate of the density function $f$,  its prediction for all the rest
of the order parameters, such as the number density, are fair
estimations of the corresponding analytic expressions in the mentioned
references.  It should be mentioned that the advantage of our method,
ultimately, is in its performance ability in situations where the simplifying assumptions are not valid and no analytic solution exists.

\begin{figure}
\centering
\includegraphics[width=0.49\columnwidth]{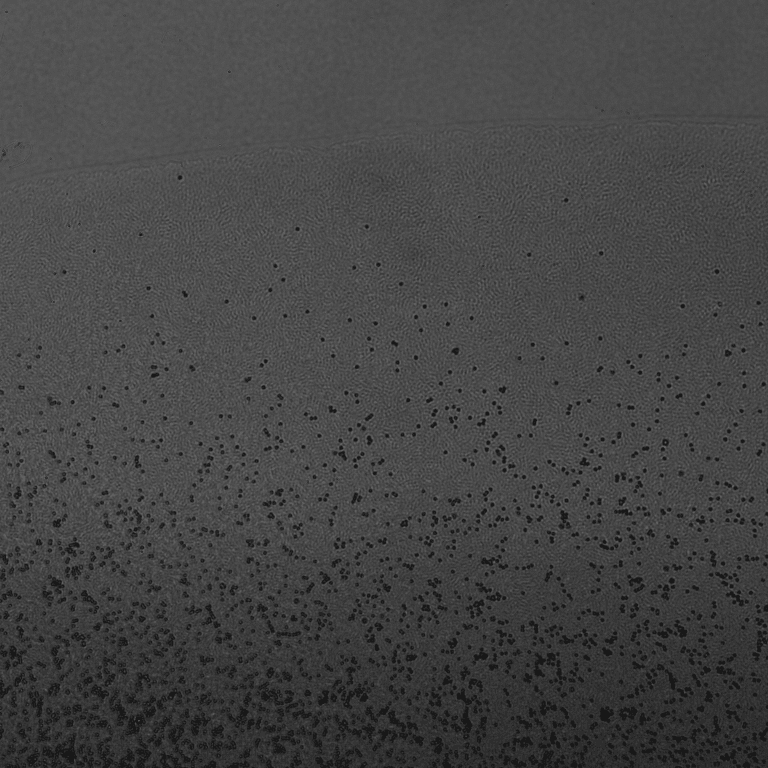}
\includegraphics[width=0.49\columnwidth]{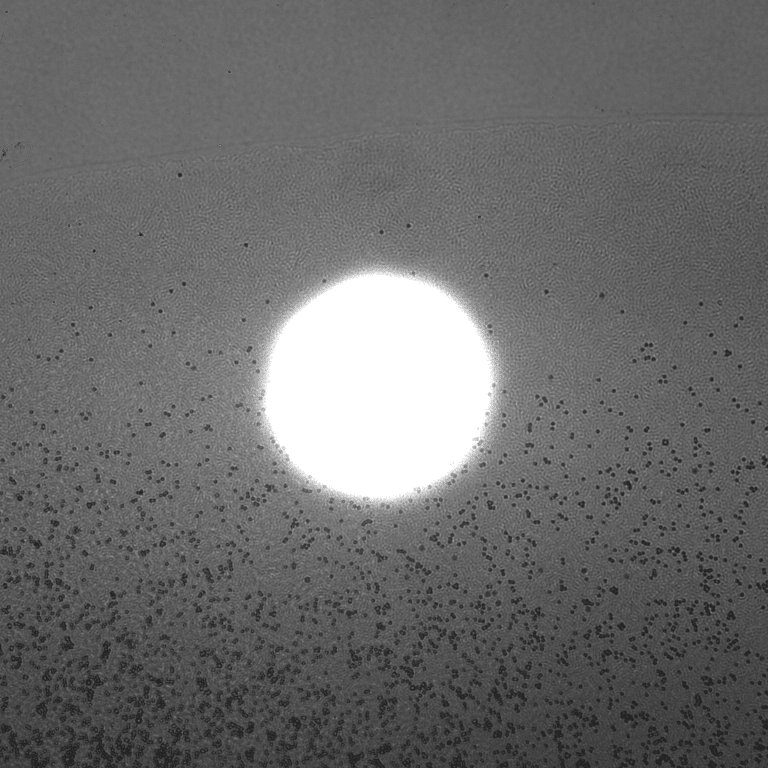}
\caption{Experimental system: Stages I and II of the passive particles actively driven by the swarm
  of bacteria. The left panel shows an early instance of the
  experiment. The right panel shows the system after ultraviolet
  light, indicated by the bright circular region, is emitted. The grey
  background shows the active bacteria while the black point-like
  particles are the passive particles.  \lb{Fig:vid19_tracer_Stages_I_II}}
\end{figure}
\section{Experimental test case: Actively-driven spherical particles}
\lb{Sec:Data}
We now turn to apply our approach to an experimental test case.  Bacterial
swarms~\cite{Kearns_2010} are a staple experimental system for active matter ~\cite{Ramaswamy_2010,Vicsek_2012}.
Each bacterium is self-propelled via molecular motors controlling
flagella to move up to speeds of tens of microns per second. While even
the motion of an individual bacterium is nontrivial, the collective motion of many bacteria exhibits a range of dynamic phenomena from jets to whirls to turbulence~\cite{Darnton_2010,Dunkel_2013}.  The predominant theoretical approach is
nematic hydrodynamics and it does predict the onset of bacterial turbulence
with associated topological
defects~\cite{Wensink_2012,Bratanov_2015}. So the collective motion of
a bacteria swarm is highly non-trivial as well.  In
addition to quantifying the motion of the swarm, researchers have
also studied particles embedded in the swarm. So while the particles themselves are
inherently passive, they are actively driven~\cite{Patteson_2019} by a
complex field.  See the
Appendix for details on the experimental set-up and conditions. 

The experiment at hand probes the effects of localized UV light on the
collective motion of the swarm and, thus, the spherical particles
driven by the swarm.  The
perturbation of the localized UV light causes the bacteria where the light
is localized to ultimately stop moving after some time window \cite{Patteson_2019}.  What
happens prior to this jamming was not resolved by the
prior experiments in terms of determining if some of the bacteria flee the region of
localized light before others eventually get jammed.  We address the issue here. 

To begin, we choose 150 consecutive movie frames well before and another 150
consecutive movie frames well after the ultraviolet light is
emitted. We refer to these two movie sets, each approximately three seconds, as stage-I and stage-II in
the following. A snapshot of the two stages can be seen in Fig.~\ref{Fig:vid19_tracer_Stages_I_II}. 
We use the trackpy package \cite{dan_allan_2019_3492186} in python to
trace the position of spherical particles over time. We apply conventional selections to remove the spurious particles.
After the positions of particles are quantified in all of the movie frames, we use the change in the positions in the consecutive frames to find the velocity of the particles.

Having reconstructed the positions and the velocities of the spherical
particles, we implement our method to estimate the phase-space density for each movie frame. A comparison of the densities reveals that they are close to the steady-state in both stage-I and stage-II. However, the system goes out of equilibrium when the ultraviolet light is turned and migrates from stage-I to stage-II. 
\begin{figure*}[thb]
\centering
\subfigure[]{
\includegraphics[width=\columnwidth]{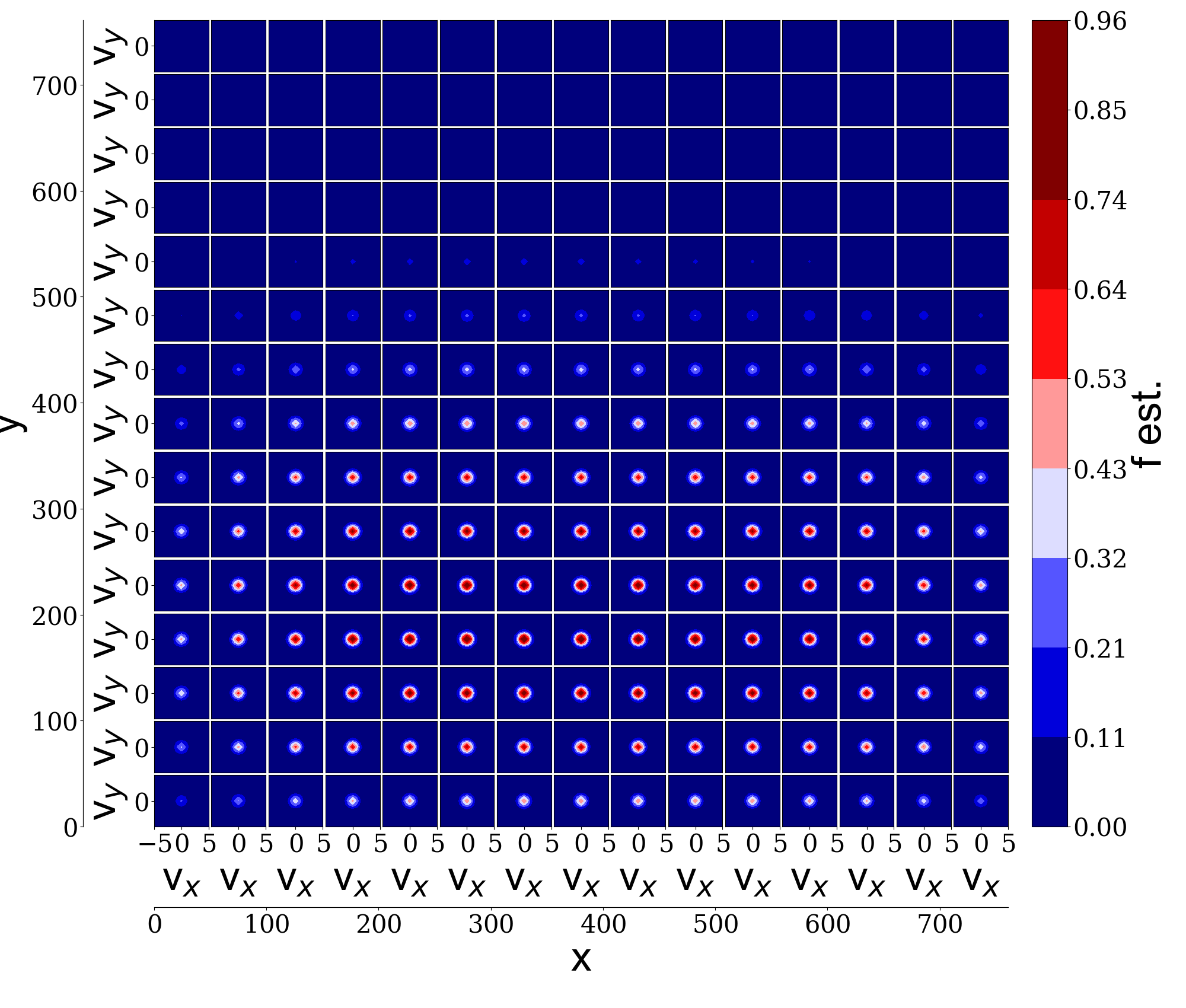}}
\subfigure[]{
\includegraphics[width=\columnwidth]{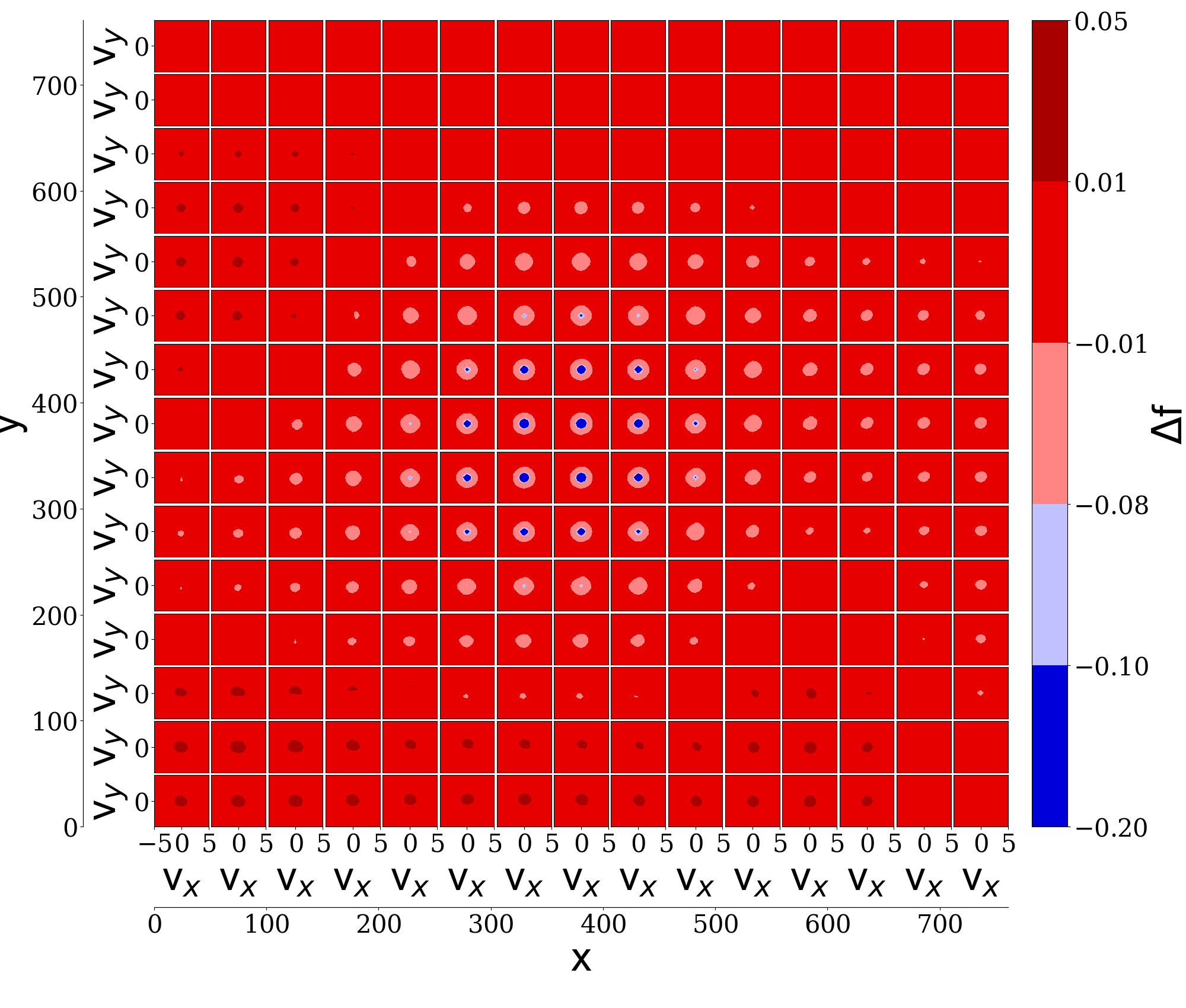}}
\subfigure[]{
\includegraphics[width=\columnwidth]{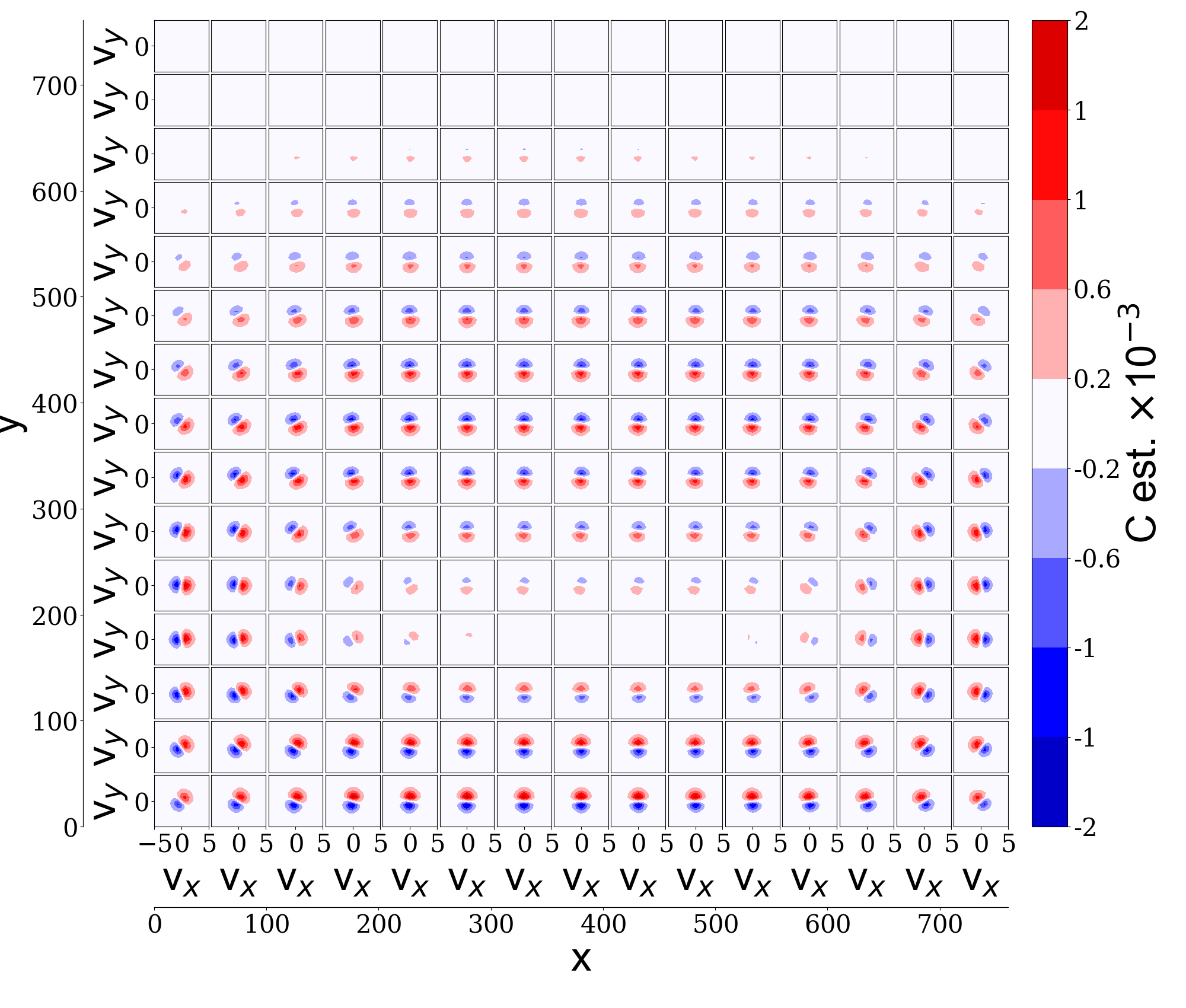}}
\subfigure[]{
\includegraphics[width=\columnwidth]{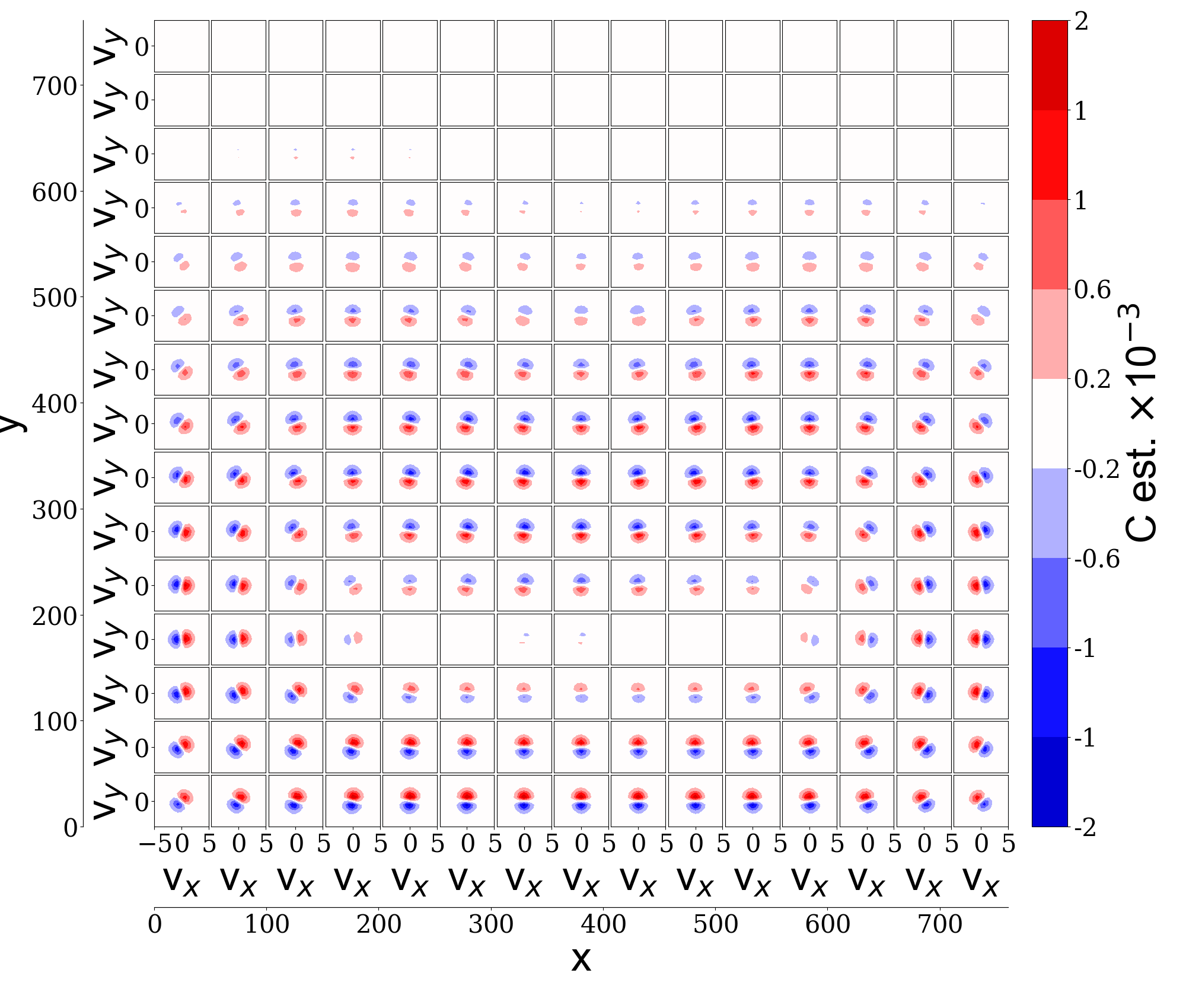}}
\caption{Experimental system: Four-dimensional presentation of quantities associated with the data, where the outer axes refer to the real-space positions, and the inner axes of the small sub-plots refer to the velocity spaces.
(a) An estimation for the phase-space density of the system in stage-I.
(b) The difference between the phase-space density of stage-II with
respect to that of stage-I. A significant drop can be seen at the
center. A mild drop in the phase-space density is shown around the
center.  An increase in the phase-space density at the bottom of the
container and at the top-left corner can be seen. This result indicates
that spherical particles are migrating away from the center in the real-space.
(c) Four-dimensional presentation of the right-hand side of the Boltzmann equation before UV light.
(d) Four-dimensional presentation of the right-hand side of the Boltzmann equation after UV light.
 \lb{Fig:vid19_tracer_frames0_150_850_1000_f} \lb{Fig:vid19_tracer_frames0_150_850_1000_C}}
\end{figure*}
Our estimations for the phase-space densities and the interaction
terms in stage-I and stage-II are shown in
Figs.~\ref{Fig:vid19_tracer_frames0_150_850_1000_f}(a,b). The figure shows
that the probability of finding particles at the center is reduced but
increased at the upper-left corner and the bottom of the
container. This figure qualitatively indicates that the spherical particles are running away from the ultraviolet light into the upper-left and bottom regions of the real-space. 
Since the spherical particles are passive, and assuming that they are
indicator of the active bacteria around them,  we argue that
the living matter---the bacteria---is evading the UV light.  Given the
crowding effect, it is difficult for all of the bacteria to flee the
UV light, just as it is in a crowd in a panic trying to exit a
stadium, for example.  However, as the bacteria become jammed in the
region of the UV light~\cite{Patteson_2019}, we speculate bacteria just outside the region
re-route since their fellow jammed bacteria in the illuminated region
now emerge as an obstacle.  It is in this re-routing that presumably directs both the
bacteria and spherical particles away from the UV light.  It would be
interesting to determine whether or not this phenomenon is robust, i.e., extends beyond our test case. 

\begin{figure*}
\centering
\subfigure[]{
\includegraphics[width=\columnwidth]{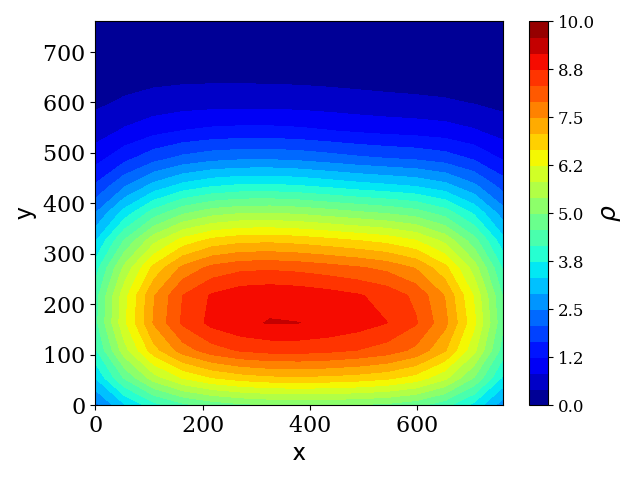}}
\subfigure[]{
\includegraphics[width=\columnwidth]{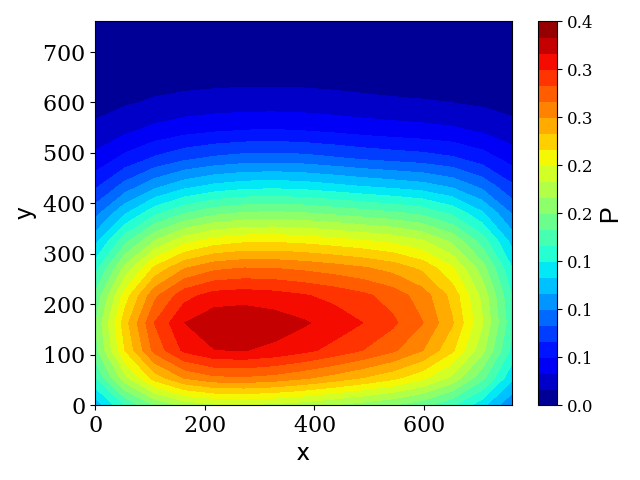}}
\subfigure[]{
\includegraphics[width=\columnwidth]{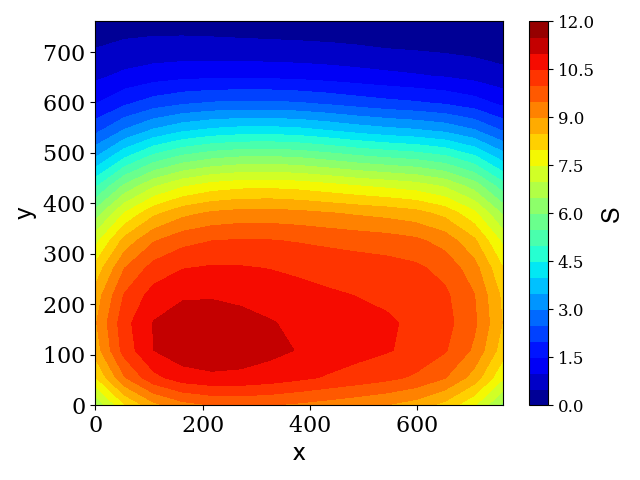}}
\subfigure[]{
\includegraphics[width=0.9\columnwidth]{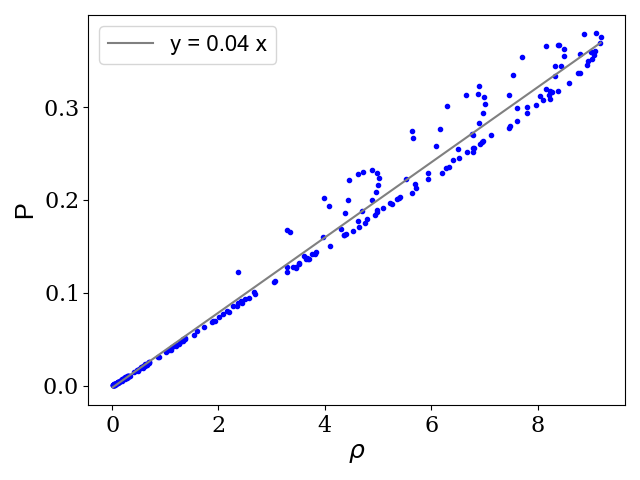}
}
\caption{Experimental system: (a) Real-space density, (b) the pressure, and (c) the entropy
of actively-driven spherical particles, reconstructed from the mean of our estimations for all of the system's snapshots in stage I.  \lb{Fig:mean_rho}
(d) Pressure versus the real-space density of the spherical particles in stage I. Comparing this plot with Fig.~\ref{Fig:TracerEqState2} indicates that the equation of state is the same in both stages I and II. 
}
\end{figure*}

\begin{figure*}[tbh]
\centering
\subfigure[]{
\includegraphics[width=\columnwidth]{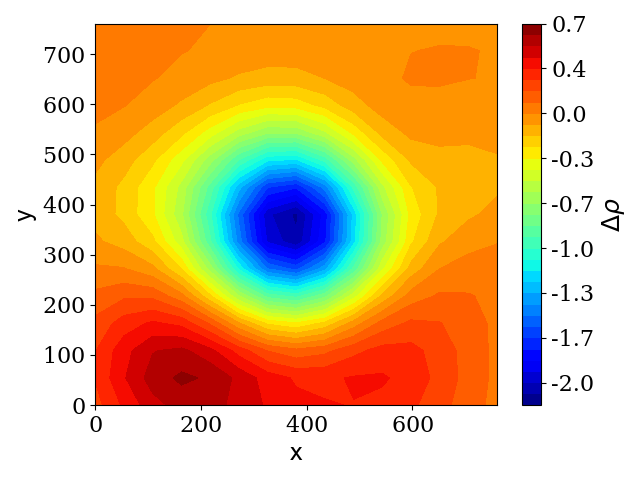}}
\subfigure[]{
\includegraphics[width=\columnwidth]{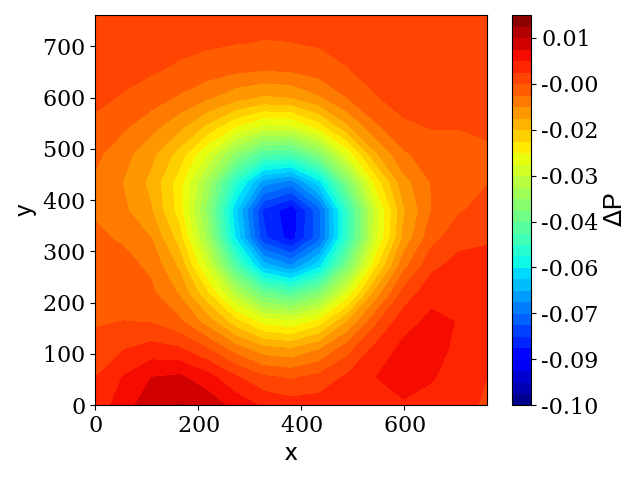}}
\caption{Experimental system: (a) The difference between the
  real-space density of the spherical particles in stages I and II. The figure quantifies the migration of the particles away from the ultraviolet light. \lb{Fig:diff_rho}
(b) The difference between the pressure of the spherical particles in
stages I and II. The figure shows that ultraviolet light leads to a
drop in the pressure in the illumination region. \lb{Fig:diff_P}
}
\end{figure*}

With this theoretical framework, we can do more. We use Eq.~\eqref{Eq:BoltzmannEq} to estimate the four-dimensional interaction term $C$  in both stages I and II, which are shown in Figs.~\ref{Fig:vid19_tracer_frames0_150_850_1000_C}(c,d). 
It should be noted that in the Boltzmann form of Eq.~\eqref{Eq:C_Boltzmann},  the interaction term has the following general form
\bqn
\lb{Eq:CFromAmplitude}
C[\vec{x},\vec{v}] = \int  d^2v'
\Big(g(\vec{x},\vec{v}\,' , \vec{v}) -g(\vec{x},\vec{v},\vec{v}\,') \Big),\nb\\
\eqn
where $g(\vec{x},\vec{v}\,' , \vec{v}) \equiv \big|{\cal{M}}\left(\vec{x},\vec{v}\,' \rightarrow \vec{v}\right)\big| f(\vec{x},\vec{v}\,')$, and ${\cal{M}}\left(\vec{x},\vec{v}\,' \rightarrow \vec{v}\right)$ is the probability amplitude of scattering one particle with velocity $\vec{v}\,'$ into velocity $\vec{v}$ off the rest of the system at position $\vec{x}$. The right hand side of the latter formulation suggests that $\int d^2v C[\vec{x},\vec{v}] =0$. This is consistent with our estimation in Fig.~\ref{Fig:vid19_tracer_frames0_150_850_1000_C} where each sub-plot is symmetric between positive and negative values with a sum of approximately zero. Since we did not enforce Eq.~\eqref{Eq:CFromAmplitude} in our estimation of $C$, the latter is a validation of our method.

Figure~\ref{Fig:mean_rho} shows the estimated number density, the
pressure, the entropy, and the equation of the state of the spherical particles in stage-I.  
These quantities are the averages of the corresponding ones overall
the movie frames of stage-I since as we discussed above,  the
phase-space densities associated with the frames in stage-I do not
change significantly over time suggesting that the system is close to
steady-state.  The latter is consistent with the linear relationship
between the number density and the pressure of the system, as can be
seen in the figure above.  Both of these findings are nontrivial given
the complexity of the bacterial swarm driving the spherical
particles.

To further quantify the migration of the spherical particles, we compute the
real-space density of particles in stage-II and show
its difference with respect to the number density in stage-I as
$\delta\rho \equiv \rho_{\text{II}}(\vec{x})-\rho_{\text{I}}(\vec{x})$
in Fig.~\ref{Fig:diff_rho}(a). The error associated with this
difference is presented in Fig.~\ref{Fig:diff_rho_Error}, which
shows that the difference in the number densities is statistically
significant. Fig.~\ref{Fig:diff_rho}(b) shows the difference in the
pressures of the spherical particles in stage-I and stage-II. As can be seen,
the density is increased near the walls of the container and decreased
at the ultraviolet site. We emphasize that our estimation of the
phase-space density, and hence the real-space density, do not solely
rely on counting the particles at the position of interest. Instead,
we take the distance of every observed particle from the position of
interest to estimate the probability of finding particles at that
point. Therefore, our estimation of the probability at a position may
be different from zero while no particle has been observed at that
point in the dataset. This point has been made clear in
Fig.~\ref{Fig:Sim4DplotsBacteria}, where using our
knowledge of underlying distribution in simulations, we have shown that the true probability at a
given point may not be zero but still, no particles exist at that
point in the dataset due to the lack of statistics. On the other hand,
in Fig.~\ref{Fig:f_comparisons1}(d), we have shown that our estimation
of the same probability is fairly close to the true one and overcomes
the lack of statistics in the dataset.

As can be seen in
Figs.~\ref{Fig:vid19_tracer_frames0_150_850_1000_C}(c,d), the
interactions in the velocity sub-spaces around the center of the
container are slightly changed due to the ultraviolet light being
shined at that location. The change in the interactions is a result of
the ultraviolet light leading to a drop in the pressure of the spherical 
particles, as can be seen in Fig.~\ref{Fig:diff_P}. On the other
hand, the system has reached a steady-state, therefore, the balance of the
forces should be zero at every location of the container. So the only
existing force on the particles that balances the pressure is their
interactions with the rest of the system, i.e. their interactions with
each other via the bacteria. 
Therefore, we can use UV light as a means of controlling the net flow of the spherical particles, and perhaps the bacteria.  By shining it at a high-density region such as the bottom of the container, we can drive the particles to move toward the empty regions at the top of the box.

It should be noted that to derive the equation of state of the
spherical particles, we have used Eqs.~(\ref{Eq:densityAndVelocity}, \ref{Eq:StressTensor}), to compute and compare the real-space
densities and the pressures of the spherical particles at every position of the container. Our best fit lines, which can be seen in Figs.~\ref{Fig:mean_rho} and \ref{Fig:TracerEqState2}, indicate that the equation of state is the same in both stages and reads
\bqn
P \simeq 0.04 \rho,
\eqn
where the coefficient indicates the effective temperature of the spherical particles in natural units. 
Here, we presented a novel derivation of the linear equation of state that is often assumed for such systems.  See for example Ref.~\cite{Copenhagen2021}.

At this point, we turn our attention to the statistical field theory
of the actively-driven spherical particles. First, we analyze the 150 consecutive movie frames of stage-I for this purpose. First, we define $\varphi(\vec{x}) \equiv \rho(\vec{x}) - \langle \rho(\vec{x})\rangle$, where the expectation symbol refers to the mean of the real-space particle densities in stage-I. Second, is the perturbative formalism via the Greens' function method
valid for this system over this time interval? To answer this
question, we look at the expectation values of the terms that might appear in the exponential of Eq.~\eqref{Eq:ProbabilityCloseToEquil} and determine their strengths. We report that $\langle \varphi(\vec{x})^2\rangle$ is the most significant term and all the other terms,  especially $\langle \partial^2 \varphi \rangle$, are at least two orders of magnitude smaller.  The leading perturbation terms are  $\langle \varphi \partial_x \varphi \rangle$, $\langle \varphi \partial_y \varphi \rangle$, and $\langle  \varphi^3 \rangle$ which are two orders of magnitude smaller than the leading term. In the following,  we neglect the rest of the corrections to the partition function. The expectation values above are shown in Fig.~\ref{Fig:vid19_tracer_frames_750_850mean_phi2}.

At this point, we construct the Green's function of the
driven spherical particle system. In light of our evaluations of the expectation values above, we can safely assume that to the leading order $Z[J] \simeq Z_0[J]$. Therefore, from Eqs.~(\ref{Eq:CorrFunction},\ref{Eq:Z_0J}), we can conclude that 
\bqn
\Delta\left(\vec{x}_1 - \vec{x}_2\right) \simeq - \langle \varphi(\vec{x}_1)\varphi(\vec{x}_2)\rangle.
\eqn
An assessment of the four-dimensional expectation value on the right-hand side of the equation above, whose value is known from data, shows that it can be split as  
\bqn
\langle \varphi(\vec{x}_1)\varphi(\vec{x}_2)\rangle \equiv
\langle \varphi(\vec{x})^2\rangle \delta(\vec{x}_1 - \vec{x}_2)
+
d(\vec{x}_1 - \vec{x}_2),
\eqn 
where $d(\vec{x}_1 - \vec{x}_2) \ll 1$, and $\langle \varphi(\vec{x})^2\rangle$  is the variance of $\varphi$ at position $\vec{x}$ regardless of other positions, and its numerical values at different $\vec{x}$ is shown in Fig.~\ref{Fig:vid19_tracer_frames_750_850mean_phi2}. 
A direct substitution of the two equations above into Eq.~\eqref{Eq:GreenFunctionEquation} proves that 
\bqn
\frac{\delta^2 S_{\text{total}}}{\delta \varphi(x_1)\delta \varphi(x_2)}\Big|_{\varphi_{_{\text{SP}}}}
\simeq 
\frac{1}{\langle \varphi(\vec{x_1})^2\rangle} \delta(\vec{x}_1 - \vec{x}_2).
\eqn
The effective free energy of the system therefore reads
\bqn
F \simeq
\frac{1}{2}\int d^2x
\frac{1}{\langle \varphi(\vec{x})^2\rangle} 
\varphi(\vec{x})^2,
\eqn
and the effective chemical potential of the system is equal to
\bqn
\mu \equiv \frac{\delta F}{\delta \varphi(x)} = 
\frac{1}{\langle \varphi(\vec{x})^2\rangle} 
\varphi(\vec{x}).
\eqn
The driven spherical particle system is, therefore, described by a Gaussian field theory
with a spatially-dependent ``mass'' and, hence, a spatially varying
chemical potential.  We have, therefore, computed the pressure, the temperature, the entropy, the effective free energy, and the effective
chemical potential of an actively-driven tracer particle collective.
We must again emphasize that the bacterial swarm is quite dynamical
and complicated at the microscopic scale,  which indicates that its
tracers are also dynamical at the same scale, though exhibiting a
different symmetry in terms of shape.  Given the complex flow of
the bacterial swarm, the novelty of our method is that it provides a simple field theoretic description of the system at the continuum scale for the tracer particles.

\section{Discussion}
\lb{Sec:Conclusion}
We have presented a data-driven approach to obtain the single-particle phase-space 
density as the solution to the stochastic dynamic equation of an
active matter system, from which physical quantities such 
as the number density, the bulk velocity, the stress tensor, the polarization vector, the nematic tensor, and entropy can be extracted.  
We do not assume a particular form for the particle interactions beforehand.  In other words, we pose an inverse method in which the data reveals the solution. 

Should a stationary state exist, an analytic field theory can be constructed to make further predictions.  In this paper, we have focused on a scalar field theory given that we were analyzing spherically-shaped particles.  
Our approach can be readily extended to nematic or polar particles,
provided the orientation and polarization data is available.  This would be interesting to do so since there
are a number of nematic hydrodynamic theories addressing the onset of
bacterial turbulence and the emergence of topological
defects~\cite{Wensink_2012,Bratanov_2015}.  Interestingly, there
exists a recent data-driven approach to quantitatively model bacterial
swarms rooted in simultaneous measurements of the orientation and
velocity fields to obtain effective parameters that can then serve as
inputs into an equation of motion with an assumed form~\cite{Li_2018}.
Again, here, we make no assumptions about the form of the solution.

In the experiment, the spherical particles were
driven by a bacterial swarm whose motility was affected by UV light.
We found that the system was in the steady-state before and after the
introduction of a localized region of UV light and found that the
particles obey a Gaussian field theory with a spatially-varying
variance, or mass, in the language of conventional field theory. 
We obtained a simple equation of state in which the pressure is
proportional to the density so that an effective temperature is
readily identifiable, despite the particles being driven by a complex
bacterial flow. We also have found that within the swarm, the spherical particles flew away from the region of UV light. We propose that this outflow
is due to re-routing of some of the {\it Serratia} since those within
the localized region of the light become jammed, along with any
spherical particles in the region as well.  
The novelty of our method is in constructing a simple continuum 
description of a system that is assumed to be complicated at the microscopic level.

In the active matter community, there has been key work addressing the
existence of an equation of state~\cite{Solon_2015,Ginot_2015}.  Specifically,
when considering active fluids, one may decide to define pressure as
the mean force per unit area exerted by the particles on a confining
wall. Alternatively, one can define extract pressure from the trace of
a bulk stress tensor, as is done here. The two definitions may not be
equivalent, at least not in the generic case, given that active matter is a
non-equilibrium phenomenon.  Some researchers have argued that an equation of state
with pressure as a state function is useless, unless one knows the
specifics of the particle-wall interactions in steady-state~\cite{Solon_2015}.  We are
able to bypass such a quandary, at least in
steady-state, because various physical quantities are extracted from
the data thereby containing all the information about the interactions
between the particles and between the particles and their confinement.
We allow the data to talk----walls and all. 
We note that recent work extracting the
single-particle density from data for equilibrium systems has been used to directly construct a free
energy~\cite{Yatsyshin_2020}.  What we present here is more general.

While we have focused on an active matter system with particles at the
micron scale, we can readily apply our method to particles at the
molecular scale. Consider, for example, an enzyme interacting with
DNA. Our method also applies to systems consisting of much larger scales, such as astronomical
systems, where such data-driven methods have been of interest~\cite{2009MNRAS.393..703M}.  Herein lies the power of
the single-particle phase-space density and its dynamic equation and its applicability to {\it any} many-body physical system, regardless of scale.  Moreover, while in this paper we have used the method on a fixed number of particles, the $C$ term in Eq.~\eqref{Eq:BoltzmannEq} also takes into account matter creation and/or annihilation, such as cell birth/death, or protein synthesis/degradation, depending on the system of focus.

Finally, our data-driven approach can be contrasted with machine-learning methods, which are typically devoid of physical
principles. It may, therefore, be tricky to use machine-learning methods to answer a physics question seeking to answer how a phenomenon occurs as
compared to a classification question. Such questions can  
ultimately be framed as a question with a yes or no answer, such an
image looking more like a cat or a dog, or a many-particle system looking more like a gas 
or a liquid. However, some scientists appreciate this point and are trying to integrate physics-based
modeling with convention machine-learning
methods~\cite{Willard_2020}. This approach could ultimately prove
fruitful. However, in this paper,  we have been guided by the single-particle phase-space density,
the stochastic dynamic equation, and its data-driven solutions in a given
situation.  By studying the system in different situations and under various perturbations, we can discover the system's underlying physics in non-equilibrium, which is key for
living matter. Moreover, by using the data to find solutions to
dynamical stochastic equations, should the form of the equations
themselves evolve with time as the system adapts, our
method can account for such adaptations---a hallmark
property of living matter.

%
%
%

\clearpage
\appendix*

\section*{Appendix A: Experimental Methods and Materials}
\renewcommand{\theequation}{A.\arabic{equation}} \setcounter{equation}{0} \renewcommand\thefigure{A.\arabic{figure}}\setcounter{figure}{0}
\lb{App:ExperimentalMeth}
{\it Serratia marcesens} (strain ATCC 274, Manassas, VA) were grown on agar
substrates prepared by dissolving 1wt\% Bacto Tryptone, 0.5 wt\% yeast
extract, 0.5 wt\% NaCl and 0.6wt\% Bacto Agar in deionized
water. Melted agar was poured into petri dishes, and 2 wt\% of glucose
solution (25 wt\%) was then added. Bacteria were then inoculated on
solidified agar plates and incubated at 34 $^oC$. Colonies formed and
grew outward from the inoculation site. 

As for the 2 micron polystyrene particles, they are cleaned by
centrifugation and then suspended in a buffer solution (67 mM NaCl
aqueous solution) with a small amount of surfactant (Tween 20, 0.03\%
by volume). A small aliquot of this particle solution (20 $\mu$L) is
gently pipetted into the bacterial colony where the 
expanding colony front meets the agar. We did not notice any visible
change in the bacteria swarm due to the addition of the spherical
particles. 

The swarm with the spherical polystyrene particles were imaged 12 – 16 hours
after inoculation. We used an inverted microscope (Nikon, Eclipse
Ti-U) to image bacteria with the free surface facing down. Depending
on the experiment either a Nikon 10$\times$  (NA $=$ 0.3) or a Nikon
20$\times$  (NA $=$ 0.45) objective was used. Images were gathered at 30 frames per
second (fps) with a Sony XCD-SX90 camera or at 60 fps and 125 fps with
a Photron Fastcam SA1.1 camera. 

To mimic the exposure of bacteria to naturally occurring wide-spectrum
light, we used a mercury vapour lamp as our light source. Standard
epifluorescence optical components with the optical light path passing
through the objective were used to focus the light on the swarm. The
bare unfiltered maximum intensity (measured at 535 nm) of the lamp was
reduced to lower, filtered (maximum) intensities using graded neutral
density filters. Since both maximum intensites depend on the
objective, care was taken to calibrate the incident intensity
carefully. Using a spectrophotometer (Thorlabs, PM100D), we measured
maximum intensities, with an unfiltered maximum intensity of 980 mW
cm$^{-2}$ (at 535 nm) for the 10$\times$ and a filtered maximum
intensity of 3100 mW cm$^{-2}$ (at 535 nm) for the 20$\times$ objectives. Variations in the
intensity of incident light were small, as indicated by checking the incident intensity on a blank slide.

\section*{Appendix B: Additional Figures}
\renewcommand{\theequation}{B.\arabic{equation}} \setcounter{equation}{0}
\renewcommand\thefigure{B.\arabic{figure}}\setcounter{figure}{0}
\lb{App:Figs}
\begin{figure*}
\centering
\includegraphics[width=\columnwidth]{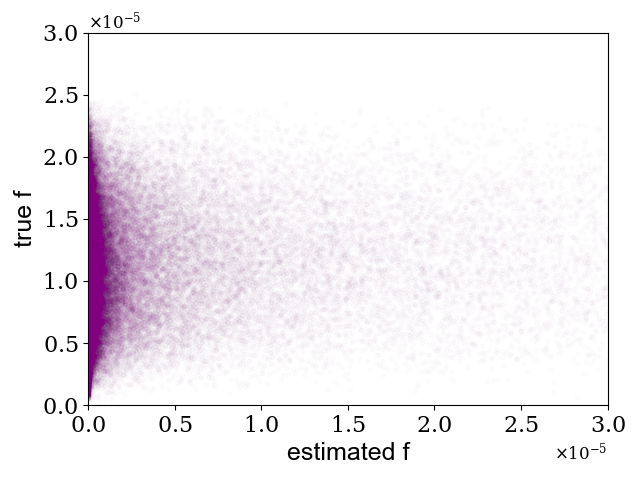}
\caption{The values of the phase-space density over the entire five-dimensions. The y-axis shows the true values and the x-axis shows the corresponding estimated values for the choice of $\rf=0.1$. 
Comparison of this figure with Fig.~\ref{Fig:f_comparisons1} indicates
the importance of tuning $\rf$ to obtain a more accurate estimation. \lb{Fig:f_comparisons2}}
\end{figure*}
\begin{figure*}
\centering
\includegraphics[width=\columnwidth]{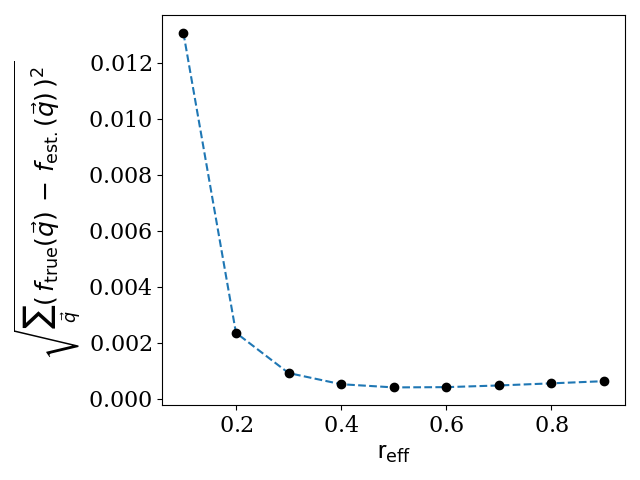}
\caption{Simulated system: The error in the y-axis defined as the square root of the sum of the square of differences between the estimated and the true phase-space density, where the sum runs over all of the points in the phase-space. The x-axis shows the chosen value of $\rf$ for the estimation. The plot indicates that $\rf \simeq 0.5$ returns the most accurate estimation. \lb{Fig:SSS_dEff}}
\end{figure*}
\begin{figure*}
\centering
\includegraphics[width=\columnwidth]{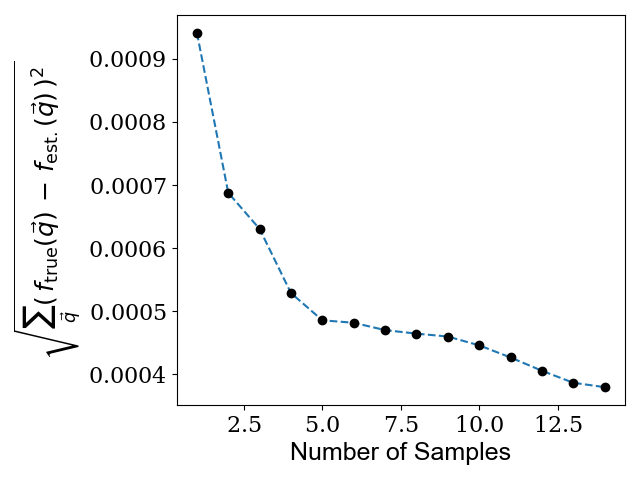}
\caption{Simulated system: Error of our estimation decreases with increasing the number of samples.\lb{Fig:SSS_NSample}}
\end{figure*}




\begin{figure*}
\centering
\includegraphics[width=\columnwidth]{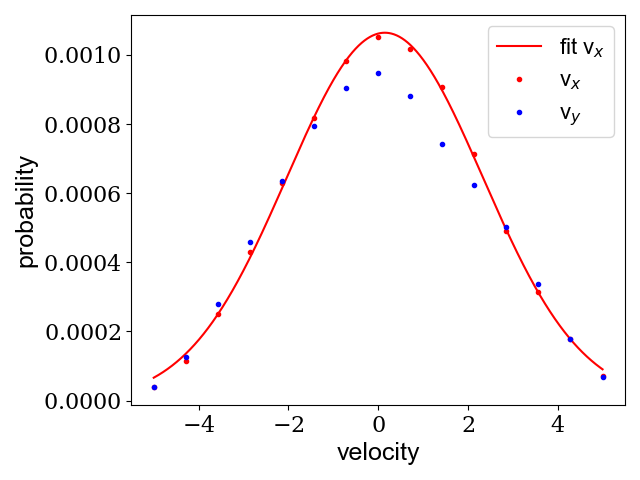}
\caption{Experimental system: The velocity distribution along either
  the $v_x$ or $v_y$ axis of
  the actively-driven particles at a fixed position. The line represents a
  Gaussian function with fitted parameters. It can be seen that the
  velocity distributions along the two axes are not the same but are
  both 
  Gaussians. Therefore, the distribution is slightly different than the
  Maxwell-Boltzmann distribution due to existence of cross-terms. \lb{Fig:IsVdistMaxwellian}}
\end{figure*}



\begin{figure*}
\centering
\includegraphics[width=\columnwidth]{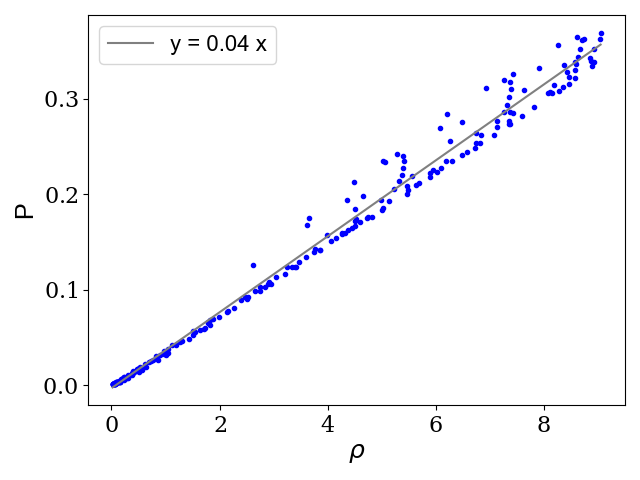}
\caption{Experimental system: Pressure versus the real-space density of the spherical particles in stateg II.  \lb{Fig:TracerEqState2}}
\end{figure*}

\begin{figure*}
\centering
\subfigure[]{
\includegraphics[width=\columnwidth]{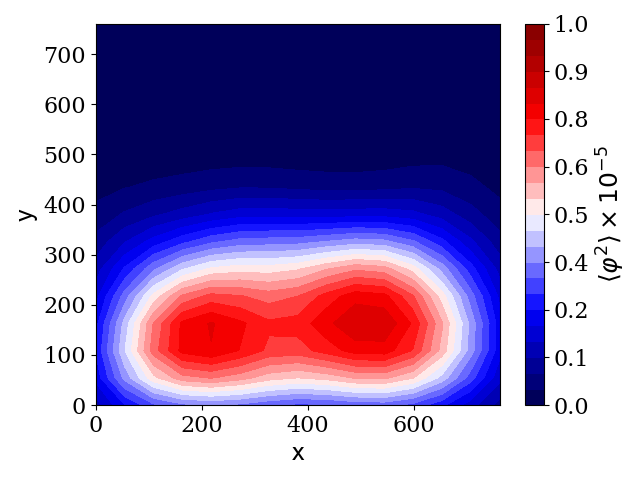}}
\subfigure[]{
\includegraphics[width=\columnwidth]{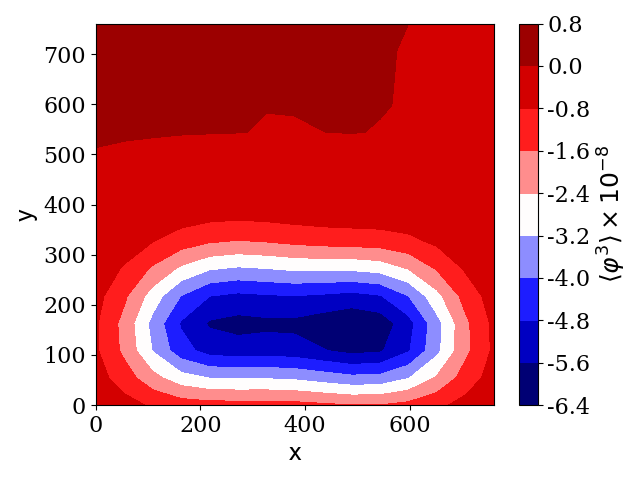}}
\subfigure[]{
\includegraphics[width=\columnwidth]{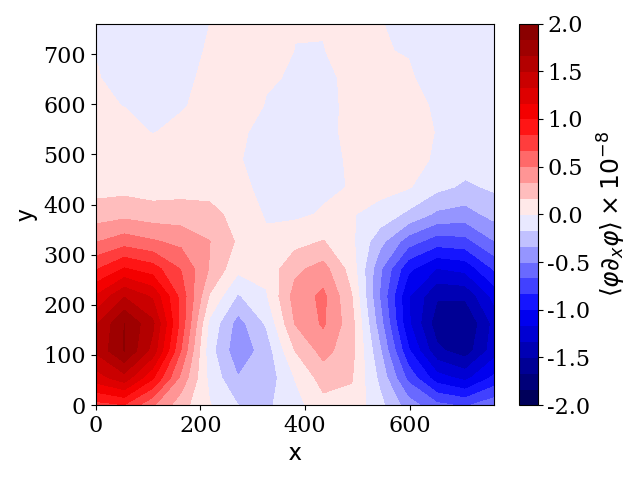}}
\subfigure[]{
\includegraphics[width=\columnwidth]{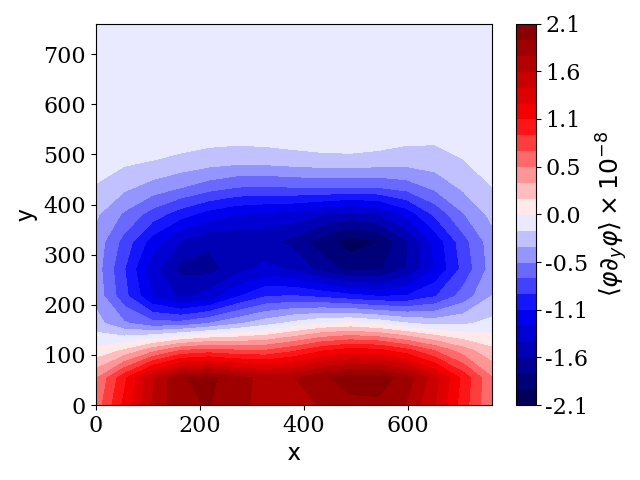}}
\caption{Experimental system: The most significant expectation values
  of powers of $\varphi \equiv  \rho - \langle \rho \rangle $ computed
  by observing the fluctuations of the quantity in stage I. These plots indicate that a perturbative description of the system is valid. \lb{Fig:vid19_tracer_frames_750_850mean_phi2}}
\end{figure*}

\begin{figure}[t]
\centering
\includegraphics[width=\columnwidth]{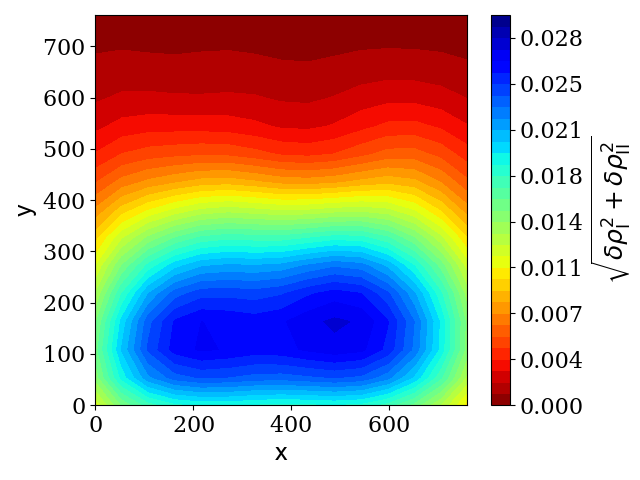}
\caption{Experimental system: The propagated error of our estimation of $\Delta\rho$ presented in Fig.~\ref{Fig:diff_rho}. The standard error of the mean in each of stages I and II is being considered as the error of our estimation $\delta \rho$.  This figure shows that our estimation of  $\Delta\rho$ is statistically significant. \lb{Fig:diff_rho_Error}}
\end{figure}

\clearpage
\begin{acknowledgments}
JMS acknowledges financial support from NSF-DMR-1832202 and an Isaac
Newton Award from the DoD.  AEP acknowledges financial support from
NSF-DEB-2033942. 
\end{acknowledgments}

\bibliography{Refs}

\end{document}